\documentclass{amsart}

\usepackage[dvips]{graphicx}
\usepackage{amsmath}
\usepackage{amsbsy}
\usepackage{amssymb}
\usepackage{amsthm, amsrefs}
\usepackage{mathrsfs}

\numberwithin{equation}{section}
\theoremstyle{plain}
\newtheorem{thm}{Theorem}[section]
\newtheorem{lemma}[thm]{Lemma}
\newtheorem{prop}[thm]{Proposition}
\theoremstyle{remark}
\newtheorem*{rem}{Remark}

\DeclareMathOperator{\dist}{dist}

\DeclareMathOperator{\spec}{spec}

\newcommand{\bx}{\boldsymbol{x}}
\newcommand{\bX}{\boldsymbol{X}}
\newcommand{\by}{\boldsymbol{y}}
\newcommand{\bz}{\boldsymbol{z}}
\newcommand{\bk}{\boldsymbol{k}}
\newcommand{\bK}{\boldsymbol{K}}
\newcommand{\bp}{\boldsymbol{p}}
\newcommand{\bq}{\boldsymbol{q}}
\newcommand{\bA}{\boldsymbol{A}}
\newcommand{\bE}{\boldsymbol{E}}
\newcommand{\bB}{\boldsymbol{B}}
\newcommand{\bJ}{\boldsymbol{J}}
\newcommand{\bbf}{\boldsymbol{f}}
\newcommand{\bbg}{\boldsymbol{g}}
\newcommand{\bzeta}{\boldsymbol{\zeta}}

\newcommand{\rd}{\,\mathrm{d}}
\newcommand{\ud}{\,\mathrm{d}}
\newcommand{\RR}{\mathbb{R}}
\newcommand{\LL}{\mathbb{L}}
\newcommand{\Or}{\mathcal{O}}

\renewcommand{\Re}{\mathfrak{Re}\,}
\renewcommand{\Im}{\mathfrak{Im}\,}

\newcommand{\bvec}[1]{\boldsymbol{#1}}

\newcommand{\veps}{\varepsilon}

\newcommand{\abs}[1]{\left\vert#1\right\vert}
\newcommand{\bra}[1]{\langle#1\rvert}
\newcommand{\ket}[1]{\lvert#1\rangle}
\newcommand{\mc}[1]{\mathcal{#1}}
\newcommand{\average}[1]{\langle #1 \rangle}
\newcommand{\nn}{\nonumber}

\newcommand{\ext}{\mathrm{ext}}

\newcommand{\gs}{\mathrm{gs}}

\renewcommand{\hat}{\widehat}
\renewcommand{\tilde}{\widetilde}

\newcommand{\ie}{\textit{i.e.}}

\newcommand{\barint}{\kern4pt
\raise3.4pt\hbox{\vrule height.6pt width7pt}%
\kern-11pt 
\int}

\title[Effective Maxwell equations from TDDFT]{Effective Maxwell
  equations from time-dependent density functional theory}

\date{\today}

\author{Weinan E}

\address{Department of Mathematics and Program in Applied and
  Computational Mathematics \\
  Princeton University \\
  Princeton, NJ 08544 \\
  Beijing International Center for Mathematical Research and
  School of Mathematical Sciences\\
  Peking University\\
  Beijing\\
  weinan@math.princeton.edu}

\author{Jianfeng Lu}

\address{Department of Mathematics \\
  Courant Institute of Mathematical Sciences \\
  New York University \\
  New York, NY 10012 \\
  jianfeng@cims.nyu.edu}

\author{Xu Yang}

\address{Department of Mathematics \\
  Courant Institute of Mathematical Sciences \\
  New York University \\
  New York, NY 10012 \\
  xuyang@cims.nyu.edu}

\thanks{The work was supported in part by the NSF grant DMS-0708026,
  grant DMS-0914336, the ONR grant N00014-01-1-0674 and the DOE grant
  DE-FG02-03ER25587.}

\begin{document}

\begin{abstract}
The behavior of interacting electrons in
a perfect crystal under macroscopic external electric and magnetic
fields  is studied.
Effective Maxwell equations for the macroscopic electric and magnetic fields
are derived starting
from time-dependent density functional theory.
Effective permittivity and permeability coefficients
are obtained.
\end{abstract}

\maketitle

\section{Introduction}

This paper is a continuation of our study on the macroscopic behavior
of interacting electrons in a crystal.  In the previous paper
\cite{ELuYang:AMASES}, we studied the Bloch dynamics of a single
electron in a crystal and introduced the Bloch-Wigner transform for
studying the semi-classical limit of Schr\"odinger equation.  We also
gave a simplified derivation of the Berry curvature term in the
effective dynamics.  In this paper, we study the collective behavior
of the interacting electrons in an insulating crystal under applied
electric and magnetic fields.  We derive the effective Maxwell
equations in this case using systematic asymptotics. In particular, we
obtain the effective permittivity and permeability coefficients for
these materials.

{}From a macroscopic viewpoint, the behavior of crystals can be
characterized as follows:
\begin{enumerate}
\item Mechanically, crystals respond to applied stress by deforming
  the crystal lattice.
\item Crystals respond to applied electric and magnetic fields by
  distorting the charge-spin distribution, or by motion of free
  electrons.  This generates electro-magnetic responses.
\end{enumerate}
The mechanical and electro-magnetic responses can be coupled together,
generating piezo-electric, magnetorestrictive and ferro-elastic
effect, etc.  The main purpose of this series of work is to provide a
systematic understanding of these macroscopic phenomena and derivation
of the effective macroscopic models from ``first principles''.

As the first principle, we choose to work with the density functional
theory \cites{HohenbergKohn:64, KohnSham:65, RungeGross:84} instead of
the many-body Schr\"odinger or Dirac equations.  This is because that
density functional theory has proven to be extremely successful for
the kind of issues we are interested in, and is at the present time
the only tractable and yet reliable models for electronic matter.
Here by density functional theory, we mostly mean Kohn-Sham density
functional theory that rely on orbitals, as is done in this paper.
But occasionally we also resort to orbital-free density functional
theory, such as the Thomas-Fermi type of models, to illustrate some of
the issues.  We refer to \cites{LiebSimon:77, Lieb:81, Lieb:83,
  BenguriaBrezisLieb:81, CattoLeBrisLions:98, BlancLeBrisLions:02,
  ProdanNordlander:03, AnantharamanCances:10} for the mathematical
works done on density functional theory.  Closely related are the
works on Hartree or Hartree-Fock models, which have also been
used as the starting point for analyzing the behavior of crystals.

When the crystal is elastically deformed, continuum mechanics model
can be derived from the Cauchy-Born rule (extended to electronic
structures). This was done for the Thomas-Fermi-von~Weisz\"acker model
in \cite{BlancLeBrisLions:02}.  In a series of works by E and Lu
\cites{ELu:ARMA, ELu:CPAM, ELu:KohnSham}, the Cauchy-Born rule was
validated for nonlinear tightbinding models and Kohn-Sham density
functional theory. One of the important ingredients in these works is
the identification of sharp stability criteria when the model has
exchange-correlation energy which might be non-convex. The issue of
stability does not occur in Thomas-Fermi-von~Weisz\"acker, Hartree or
reduced Hartree-Fock model, since these models do not include
exchange-correlation energy. Further in this direction, E and Lu
studied in \cite{ELu:spinTFDW} the continuum limit of the
spin-polarized Thomas-Fermi-von~Weisz\"acker-Dirac model under
external macroscopic magnetic fields. Under stability conditions for
plasmon and magmon, a micromagnetics energy functional was derived.

One interesting by-product of the work in
\cite{ELu:KohnSham} is an effective model for the macroscopic
electric potential as a result of the crystal deformation, which
exhibits a coupling between the mechanical and electric responses.

Cances and Lewin studied the reduced Hartree-Fock model for a crystal
under a macroscopic external potential and proved that the implied
macroscopic potential satisfies an effective Poisson equation.  In
particular, they established the validity of the well-known
Adler-Wiser formula for the permittivity tensor \cite{CancesLewin:10}.

In this work,
we consider the time-dependent Kohn-Sham density functional theory in
the presence of external macroscopic electric and magnetic fields. The
questions of interest are whether macroscopic Maxwell equations that
describe the electromagnetic fields can be derived in the continuum
limit from the underlying microscopic theory, and in particular, how
to obtain effective permittivity and permeability for materials from
electronic structure models. We resolve these issues using asymptotic
analysis. To rigorously justify the asymptotic derivation, one
needs to identify correct stability conditions for time dependent models.
This will be left to future publications.

The paper is organized as follows. In Section \ref{sec:tddft}, we
introduce the time-dependent Kohn-Sham density functional theory.
Section \ref{sec:main} describes the model setup and presents the main
results. The asymptotic derivation is given in Section
\ref{sec:asymptotics}, Section \ref{sec:time} and Section
\ref{sec:freq}. We make conclusive remarks in Section
\ref{sec:conclusion}.

\section{Time-dependent density functional theory}\label{sec:tddft}

Time dependent density functional theory (TDDFT) \cite{RungeGross:84}
is an extension of (static) density functional theory to the dynamics
of interacting electrons. In TDDFT, the electron dynamics is governed
by $N$ one-electron time dependent Schr\"odinger equations with
effective one-body Hamiltonian depending on electron density and/or
electron current density.\footnote{When the effective Hamiltonian
  depends on electron current density, the model is usually called
  time dependent current density functional theory (TDCDFT)
  \cites{Vignale:95, VignaleKohn:96} in physics literature, although
  we still use the name of time dependent density functional theory in
  this paper.}

The TDDFT model takes the following form in physical units in
$\RR^3$,
\begin{align}\label{eqn:phys_1}
  & i\hbar \frac{\partial \psi_j}{\partial t} = \frac{1}{2m_e}
  \left(-i\hbar\nabla - \frac{e}{c} (\bA +
    \bA_{\ext} ) \right)^2 \psi_j + e(V + V_{\ext}) \psi_j, \\
  & -\Delta \phi = \frac{e}{\epsilon_0}(\rho-m), \\
  & \frac{1}{c}\frac{\partial}{\partial t}
  \left(\frac{1}{c}\frac{\partial}{\partial t}\bA +\nabla \phi
  \right)-\Delta \bA=\frac{e}{c\epsilon_0}\bJ, \label{eqn:phys_3} \\
  & \nabla\cdot \bA = 0, \label{eqn:cg}\\
  & V(t,\bx) = \phi(t, \bx) + \eta( \rho(t, \bx) ). \label{eqn:phys_5}
\end{align}
Here $\psi_j$, $j = 1, \ldots, N$, is the one-particle wave
function, $\bA$ is the vector potential and $\phi$ is the scalar
potential generated by electrons. The electric and magnetic fields are
given by
\begin{equation*}
 \bE=-\nabla\phi-\frac{\partial \bA}{\partial t},\qquad \bB=\nabla\times\bA.
\end{equation*}
The system is invariant under the gauge transform,
\begin{equation*}
  \bA\rightarrow\bA+\nabla\chi,
  \qquad \phi\rightarrow\phi-\frac{\partial\chi}{\partial t},
\end{equation*}
and hence we fix the Coulomb gauge \eqref{eqn:cg} in the model.
$\bA_{\ext}$ and $V_{\ext}$ are the external vector and scalar
potentials. The electron number density and electron current density
are denoted by $\rho$ and $\bJ$ respectively in the equations, and are
given in terms of $\{\psi_j\}_{j=1}^N$ by
\begin{align}
  & \rho(t,\bx) = \sum_{j=1}^N \abs{\psi_j(t,\bx)}^2, \\
  & \bJ(t,\bx) = \frac{\hbar}{m_e} \sum_{j=1}^N \Im(\psi^*_j(t,\bx)
  \nabla\psi_j(t,\bx) )-\frac{e}{m_ec} \rho(t,\bx)\bA(t,\bx).
\end{align}
The function $m(\bx)$ is the background charge density contributed
by the nuclei. We assume that the nuclei are fixed so that $m(\bx)$ is
independent of time. In \eqref{eqn:phys_1}, we have the physical
constants electron mass $m_e$, electron charge $e$, Planck constant
$\hbar$, dielectric constant in vacuum $\epsilon_0$ and speed of
light in vacuum $c$.

The electric and magnetic fields are given by vector and scalar
potentials (in the Coulomb gauge),
\begin{align}
  & E = -\nabla \phi -\frac{1}{c}\frac{\partial}{\partial t}\bA; \\
  & B = \nabla\times\bA.
\end{align}

We make some remarks about the model.
\begin{enumerate}
\item The spin is ignored in the above TDDFT model. As a result, only
  the orbital magnetization is considered, while spin magnetization is
  not present. The extension to include spin in the model is
  straightforward though.
\item We adopt the adiabatic local density approximation
  \cites{CeperleyAlder:80, ZangwillSoven:80} for the
  exchange-correlation potential, denoted as $\eta$ in equation
  \eqref{eqn:phys_5}. This means that the exchange-correlation
  potential is a function of local electron density only. No
  exchange-correlation vector potential is included in the model.
  Generally, the exchange-correlation potential can depend on the
  local electron current density and the derivatives of electron
  density and current density. Exchange-correlation vector potential
  can also be added.  The extension to these general models is in
  principle possible, but will complicate the formulations and
  derivations in the discussions below.
\item The model agrees with what physicists commonly use in
  practical applications (for instance \cite{BeIwRuYa:00}). Of
  course, whether the model gives a good prediction of the
  time-evolution of electronic structure depends on the choice of
  pseudo-potential, the choice of exchange-correlation functional,
  and sometimes requires additional terms like
  exchange-correlation vector potential. We will not go into the
  details of this discussion.
\end{enumerate}

\medskip
\noindent\textbf{Nondimensionalization and high frequency scaling}

We consider the situation when the applied external fields
to the system have a much larger characteristic length compared to
the atomistic length scale (lattice parameter). For this purpose, we
perform nondimensionalization to the set of equations and identify
small parameters.

We introduce two sets of units to rescale the system. One is the
microscopic unit in which we denote the units of time, length, mass
and charge as $[t],\;[l],\;[m],\;[e]$; the other is the macroscopic
unit in which we denote the units of time and length as $[T],\;[L]$.
It means that for example the characteristic time scale for
macroscopic fields is $[T]$, while that for microscopic fields is
$[t]$.  We will consider the macroscopic behavior of the system
under macroscopic external potentials within the high frequency
regime, in other words, the regime
\[
[T]\sim [t],\quad [L]\gg[l].
\]
The small parameter is identified as $\veps=[l]/[L]$. Physically, the
high frequency regime means that we are interested in the dynamics of
the electronic structure and the corresponding dynamics of the
electromagnetic fields on the time scale that is comparable to the
characteristic time scale of the quantum system. At longer time
scale, different physical phenomena might occur and is not covered by
the results here. In particular, this is different from
the scaling used when considering the semi-classical limit.

Using these two sets of units, we can represent all physical
constants and quantities in suitable units so that they become
nondimensional and have values of order $\Or(1)$. For example,
Planck constant, vacuum dielectric constant and speed of light can
be written as
\[\hbar=1\times\frac{[m][l]^2}{[t]},\quad
\epsilon_0=1\times\frac{[e]^2[t]^2}{[m][l]^3},\quad
c=1\times\frac{[L]}{[T]}.
\]
The temporal and spatial derivatives are rescaled as
\[
\frac{\partial}{\partial t} \longrightarrow
\frac{1}{[T]}\frac{\partial}{\partial t},\quad \nabla
\longrightarrow \frac{1}{[L]}\nabla.
\]
The physical quantities are rewritten as
\[e\bA=\tilde{A}\frac{[m][l]^2}{[t]^2},\quad eV=\tilde{V}\frac{[m][l]^2}{[t]^2},
\quad \rho=\tilde{\rho}\frac{1}{[L]^3}, \quad
\bJ=\tilde{\bJ}\frac{[l]^2}{[t][L]^4},\] where
$\tilde{A},\;\tilde{V},\;\tilde{\rho}\;,\tilde{\bJ}$ are
nondimensional quantities.

Substituting all the above into the system
\eqref{eqn:phys_1}-\eqref{eqn:phys_5} produces the
nondimensionalized TDDFT equations (the tildes are dropped for
simplicity),
\begin{align}
  & i \frac{\partial \psi_j}{\partial t} = \frac{1}{2}
  \left(-i\veps\nabla - \veps (\bA +
    \bA_{\ext} ) \right)^2 \psi_j + (V + V_{\ext}) \psi_j, \label{eqn:phys_4} \\
  & -\Delta \phi = \veps(\rho-m), \\
  & \frac{\partial^2}{\partial t^2} \bA - \Delta \bA +
  \frac{\partial}{\partial t} \nabla \phi =\veps^2 \bJ,  \\
  & \nabla\cdot \bA = 0, \\
  & V(t,\bx) = \phi(t, \bx) + \eta( \veps^3 \rho(t, \bx)
  ). \label{eqn:phys_6}
\end{align}
The density and current are given by
\begin{align}
  & \rho(t,\bx) = \sum_{j=1}^N \abs{\psi_j(t,\bx)}^2, \\
  & \bJ(t,\bx) = \veps \sum_{j=1}^N \Im\big(\psi^*_j(t,\bx)
  \nabla\psi_j(t,\bx) \big)-\veps \rho(t,\bx)\bA(t,\bx).
\end{align}

\section{The effective Maxwell equations in crystal}\label{sec:main}

\subsection{Unperturbed system}

Let $\mathbb{L}$ be a lattice with unit cell $\Gamma$. Denote the
reciprocal lattice as $\mathbb{L}^{\ast}$ and the reciprocal unit cell
as $\Gamma^{\ast}$. We consider system as a crystal $\veps
\mathbb{L}$, so that $\veps$ is the lattice constant (the micro length
scale used in the non-dimensionalization). Therefore, the charge
background is given by
\begin{equation}
  m^{\veps}(\bx) = \veps^{-3} m_0(\bx/\veps),
\end{equation}
where $m_0$ is $\Gamma$-periodic. Note that the factor $\veps^{-3}$
comes from rescaling so that the total background charge in one unit
cell is the constant $Z$ independent of $\veps$, \ie
\begin{equation}\label{eq:m_charge}
  \int_{\veps \Gamma} m^{\veps}(\bx)\ud \bx =Z.
\end{equation}
We introduce
the following notations for cell average in physical and reciprocal
spaces
\begin{equation*}
  \langle f(\bz) \rangle_{\bz}=\int_{\Gamma}f(\bz)\rd\bz, \qquad
  \barint_{\Gamma^*}g(\bk)\rd\bk=\frac{1}{\abs{\Gamma^*}}
  \int_{\Gamma^*}g(\bk)\rd\bk.
\end{equation*}
When there are no external applied potentials ($V_{\ext} = 0,
\bA_{\ext} = 0$), TDDFT system can be written as
\begin{align}
  & i\frac{\partial \psi_j^{\veps}}{\partial t}=
  \tfrac{1}{2}\bigl(-i\veps\nabla
  -\veps \bA^\veps \bigr)^2 \psi_j^{\veps} +V^\veps \psi_j^{\veps}, \\
  & -\Delta \phi^\veps = \veps\bigl(\rho^\veps(t,\bx)-m^{\veps}(\bx)
  \bigr),  \\
  & \frac{\partial^2}{\partial t^2}\bA^\veps -\Delta \bA^\veps+
  \frac{\partial}{\partial t}\left(\nabla
    \phi^\veps\right)=\veps^2\bJ^\veps, \quad
  \nabla\cdot \bA^\veps =0, \\
  & V^\veps(t, \bx) =\phi^\veps(t,\bx)+\eta(\veps^3\rho^{\veps}(t,
  \bx)).
\end{align}

We assume that there exists a ground state for the unperturbed
system, with density having the lattice periodicity
\begin{equation}
  \rho^{\veps}(\bx) = \veps^{-3}\rho_{\gs}(\bx/\veps),
\end{equation}
where $\rho_{\gs}$ $\Gamma$-periodic. Absence of external
perturbation implies that the system will stay at the ground state
with no electronic current and hence no induced vector potential,
\begin{equation*}
  \bJ^{\veps}(t, \bx) = 0, \quad \bA^{\veps}(t, \bx) = 0.
\end{equation*}
The evolution equations are then simplified as
\begin{align}\label{eqn:Schr0}
  i\frac{\partial \psi_j}{\partial t} &=
  -\frac{\veps^2}{2}\Delta\psi_j +V^\veps(\bx) \psi_j, \\
  V^\veps &=\phi^\veps(\bx)+ \eta(\rho_{\gs}(\bx/\veps)), \\
  -\Delta \phi^\veps &=\veps^{-2} (\rho_{\gs}(\bx/\veps)- m_0(\bx/\veps)).
  \label{eqn:Poisson0}
\end{align}
Note that the potential is independent of time if there is no external
perturbation. It is easy to see that the potential is
$\veps\Gamma$-periodic. We denote the potential corresponding to the
ground state as $V^{\veps}(\bx) = v_0(\bx/\veps) = v_{\gs}(\bx/\veps)$
where $v_{\gs}$ is $\Gamma$-periodic.

The Hamiltonian operator for the ground state is independent of time, given by
\begin{equation}
  H_0^{\veps} = - \frac{\veps^2}{2} \Delta + v_{\gs}(\bx/\veps).
\end{equation}
Define the rescaling operator $\delta_{\veps}$ as
\begin{equation}
  (\delta_{\veps} f)(\bx) = \veps^{-3/2} f(\bx/\veps).
\end{equation}
It is easy to check that $\delta_{\veps}$ is a unitary operator. We
have
\begin{equation}
  H_0 = -\frac{1}{2}\Delta + v_0(\bx) = \delta_{\veps}^{\ast} H_0^{\veps}
  \delta_{\veps}.
\end{equation}
Since $v_0$ is $\Gamma$-periodic, $H_0$ is invariant under the
translation with respect to the lattice $\mathbb{L}$.
The standard Bloch-Floquet theory gives the decomposition of $H_0$,
\begin{equation}
  H_0 = \barint_{\Gamma^{\ast}} H_{0,\bk} \ud \bk,
\end{equation}
where $H_{0,\bk}$ is an operator defined on $L_{\bk}^2(\Gamma)$ for
each $\bk \in \Gamma^{\ast}$,
\begin{equation*}
  L_{\bk}^2(\Gamma) = \{ f \in L^2(\Gamma) \mid
  \tau_{\bvec{R}} f = e^{- i \bvec{R} \cdot \bk} f, \ \forall \bvec{R}
  \in \mathbb{L} \}.
\end{equation*}
Here $\tau_{\bvec{R}}$ is the translation operator, \ie
$\;\tau_{\bvec{R}} f(\bx)=f(\bx+\bvec{R})$.  The operator $H_{0,\bk}$
has the spectral representation
\begin{equation}
  H_{0,\bk} = \sum_{n} E_n(\bk) \lvert \psi_{n,\bk} \rangle \langle
  \psi_{n,\bk} \rvert,
\end{equation}
where $E_n(\bk)$ is the $n$-th eigenvalue of $H_{0,\bk}$, and
$\psi_{n,\bk}$ is the corresponding eigenfunction (named as Bloch
wave in literature) with
\begin{equation*}
  u_{n,\bk}(\bx) = e^{-i\bk \bx} \psi_{n,\bk}(\bx)
\end{equation*}
being $\Gamma$-periodic. Moreover, the spectrum $\spec(H_0)$ has the
band structure,
\begin{equation*}
  \spec(H_0) = \bigcup_{n} \bigcup_{\bk\in\Gamma^{\ast}} E_n(\bk).
\end{equation*}
Denote the spectrum for the first $Z$ bands by $\sigma_Z$,
\begin{equation}
  \sigma_Z = \bigcup_{n = 1}^Z \bigcup_{\bk\in \Gamma^{\ast}} E_n(\bk),
\end{equation}
where $E_n(\bk)$ is the $n$-th eigenvalue of $H_0$.

We assume that the ground state satisfies the gap condition,
\begin{equation}
  \dist(\sigma_Z, \spec(H_0))\backslash\sigma_Z) = E_g.
\end{equation}
In physical terminology, the system is called a band insulator with
band gap $E_g$.

For convenience, we use the bra and ket notations
\begin{equation*}
  \bra{f(\bzeta)}  \mathcal{K} \ket{
    g(\bzeta)}_{L^2(\Gamma)}=\int_{\Gamma}f^*(\bzeta)\mathcal{K}g(\bzeta)\rd\bzeta,
\end{equation*}
where $\mathcal{K}: L^2(\Gamma) \to L^2(\Gamma)$ is a linear operator.

\subsection{Macroscopic perturbation}

We are interested in the dynamics of the electronic structure in the
presence of the external potentials $\bA_{\ext}(t, \bx)$ and
$V_{\ext}(t,\bx)$. We assume that $\bA_{\ext}$ and $V_{\ext}$ are
smooth functions in both $t$ and $\bx$ and periodic in space in the
domain $\Gamma$. Hence, the characteristic length scales of external
applied fields are $\Or(1)$, while the lattice constant is
$\Or(\veps)$.  We consider the continuum limit $\veps \to 0$; the
disparity of the space scales leads to macroscopic Maxwell equations.

We consider the following system with periodic conditions on
$\Gamma$,
\begin{align}\label{eqn:Schr}
  & i\frac{\partial \psi_j^{\veps}}{\partial t}=H^{\veps} \psi_j^{\veps}, \\
  & -\Delta \phi^\veps = \veps\bigl(\rho^\veps(t,\bx)-m_0(\bx/\veps)
  \bigr), \label{eqn:Poisson} \\
  & \frac{\partial^2}{\partial t^2}\bA^\veps -\Delta \bA^\veps+
  \frac{\partial}{\partial t}\left(\nabla
    \phi^\veps\right)=\veps^2\bJ^\veps,
  \label{eqn:VecPot} \\
  & \nabla\cdot \bA^\veps =0, \\
  & V^\veps(t, \bx) =\phi^\veps(t,\bx)+\eta(\veps^3\rho^{\veps}(t,
  \bx)), \label{eqn:V_tot}
\end{align}
where the Hamiltonian operator $H^{\veps}$ is given by
\[
H^{\veps} = \frac{1}{2}\bigl(-i\veps\nabla
-\veps(\bA^\veps+\bA_{\ext}) \bigr)^2 +V^\veps + V_{\ext}.
\]
We have used the superscript $\veps$ to make explicit the dependence
on the small parameter. The density and current is then given by
\begin{equation*}
  \rho^\veps=\sum_{k=1}^{Z/\veps^3}\abs{\psi_j^{\veps}}^2,\quad
  \bJ^\veps=\veps\sum_{k=1}^{Z/\veps^3}
  \Im\left((\psi_j^{\veps})^*\nabla\psi_j^{\veps}\right)-\veps\bA^{\veps}
  \rho^{\veps}.
\end{equation*}
Here $Z$ is the number of electrons in one unit cell, which equals
to the background charge \eqref{eq:m_charge}. We remark that in the
domain $\Gamma$, since the lattice constant is $\veps$, there are
$\veps^{-3}$ unit cells in total, and hence $N = Z\veps^{-3}$
electrons under consideration.

\subsection{Main result}\label{sec:mainresult}
Define the limiting macroscopic potentials as
\begin{align}
  & U_0(t, \bx) = \lim_{\veps \to 0} \bigl(V^{\veps}(t, \bx) + V_{\ext}(t,
  \bx) - v_{\gs}(\bx/\veps)\bigr), \\
  & A_0(t, \bx) = \lim_{\veps \to 0} \bigl(A^{\veps}(t, \bx) + A_{\ext}(t,
  \bx)\bigr);
\end{align}
and the corresponding electric and magnetic fields
\begin{align}
  & \bE(t, \bx) = - \nabla_{\bx} U_0(t, \bx) - \frac{\partial}{\partial t}
  \bA_0(t, \bx), \\
  & \bB(t, \bx) = \nabla_{\bx} \times \bA_0(t, \bx).
\end{align}
Define the electric field in frequency space,
\[
\hat{\bE}(\omega, \bx)=\int_0^\infty e^{i\omega t} \bE(t, \bx)\rd t,
\]
and similarly for $\hat{\bB}$, $\hat{U_0}$ and $\hat{\bA_0}$. We have
\begin{align}
  & \hat{\bE}(\omega, \bx) =
  -\nabla_{\bx}\hat{U_0}(\omega, \bx) + i\omega \hat{\bA_0}(\omega, \bx), \\
  & \hat{\bB}(\omega, \bx) = \nabla_{\bx}\times\hat{\bA_0}(\omega, \bx).
\end{align}

We will show that TDDFT system gives arise to the effective Maxwell
system as
\begin{align}\label{eqn:E}
  & \nabla_{\bx}\cdot\bigl(\mathcal{E}(\omega)\hat{\bE}(\omega,
  \bx)\bigr) = \hat{\rho_{\ext}}(\omega, \bx), \\
  & \nabla_{\bx}\cdot\hat{\bB}(\omega, \bx) = 0, \\
  & \nabla_{\bx}\times\hat{\bE}(\omega, \bx) = i\omega\hat{\bB}(\omega, \bx), \\
  & \nabla_{\bx}\times\hat{\bB}(\omega, \bx) =
  -i\omega\mathcal{E}(\omega)\hat{\bE}(\omega, \bx)
  +\hat{\bJ_{\ext}}(\omega, \bx), \label{eqn:B}
\end{align}
where
\begin{align*}
  & \hat{\rho}_{\ext}(\omega, \bx)= \int_0^\infty e^{i\omega t}
  \rho_{\ext}(t, \bx)\rd t, \\
  & \hat{\bJ}_{\ext}(\omega, \bx) = \int_0^\infty e^{i\omega t}
  \bJ_{\ext}(t, \bx)\rd t,
\end{align*}
with $\rho_{\ext}$ and $\bJ_{\ext}$ given by
\begin{align*}
  & \rho_{\ext}(t,\bx)=-\Delta_{\bx}V_{\ext}(t,\bx), \\
  & \bJ_{\ext}(t,\bx)=\frac{\partial^2}{\partial t^2}\bA_{\ext}
  -\Delta_{\bx} \bA_{\ext}+ \frac{\partial}{\partial
    t}\left(\nabla_{\bx} V_{\ext}\right).
\end{align*}

The system \eqref{eqn:E}-\eqref{eqn:B} are (nonlocal) Maxwell
equations with dynamic dielectric permittivity matrix
$\mathcal{E}_{\alpha\beta}=\delta_{\alpha\beta}+A_{\alpha\beta}$ given
by
\begin{align*}
  A_{\alpha\beta}(\omega)&  = \\
  &\sum_{n\leq
    Z}\sum_{m>Z}\barint_{\Gamma^*}\frac{1}{\omega+\omega_{mn}(\bk)}
  \overline{\langle u_{n,\bk}|i\partial_{\bk_\alpha}|u_{m,\bk}
    \rangle}_{L^2(\Gamma)}\langle
  u_{n,\bk}|i\partial_{\bk_\beta}|u_{m,\bk} \rangle_{L^2(\Gamma)}\rd\bk \\
  & -\sum_{n\leq
    Z}\sum_{m>Z}\barint_{\Gamma^*}\frac{1}{\omega-\omega_{mn}(\bk)}
  \langle u_{n,\bk}|i\partial_{\bk_\alpha}|u_{m,\bk}
  \rangle_{L^2(\Gamma)}\overline{\langle
    u_{n,\bk}|i\partial_{\bk_\beta}|u_{m,\bk} \rangle}_{L^2(\Gamma)}\rd\bk \\
  & -\frac{2i}{\omega}\Im\sum_{n\leq Z}\sum_{m>Z}\barint_{\Gamma^*}
  \overline{\bra{u_{n,\bk}}i\partial_{\bk_{\alpha}}\ket{u_{m,\bk}}}
  \bra{u_{n,\bk}}i\partial_{\bk_{\beta}}\ket{u_{m,\bk}}\rd\bk\\
  &-\bigg\langle\hat{f}_{\alpha}^*(\omega, \bz)
  \mathcal{V}\left(\mc{I} -\hat{\chi}_{\omega}
    \mathcal{V}\right)^{-1}\hat{f}_\beta(\omega, \bz)
  \bigg\rangle_{\bz}.
\end{align*}
Here the potential operator $\mc{V}$ is
the linearized effective potential operator at the equilibrium density $\rho_0$:
\begin{align*}
  & (\mc{V} f)(\bz) =\phi(\bz) +\eta'(\rho_0(\bz))f(\bz), \\
  & -\Delta_{\bz}\phi(\bz) = f(\bz), \quad \average{\phi} = 0.
\end{align*}
The operator $\hat{\chi}_{\omega}$ and the function $\hat{\bbf}$ are
defined as
\begin{align*}
  \hat{\chi}_{\omega} v = & -\sum_{n\leq
    Z}\sum_{m>Z}\barint_{\Gamma^*}\frac{1}{\omega + \omega_{mn}(\bk)}
  u_{n,\bk} u_{m,\bk}^*\langle
  u_{n,\bk}\rvert v \lvert u_{m,\bk} \rangle_{L^2(\Gamma)}\rd\bk \\
  &+\sum_{n\leq Z}\sum_{m>Z}\barint_{\Gamma^*}\frac{1}{\omega -
    \omega_{mn}(\bk)} u_{n,\bk}^* u_{m,\bk}\overline{\langle
    u_{n,\bk} \rvert v \lvert u_{m,\bk} \rangle}_{L^2(\Gamma)}\rd\bk, \\
  \hat{\bbf}(\omega)=&-\sum_{n\leq
    Z}\sum_{m>Z}\barint_{\Gamma^*}\frac{1}{\omega + \omega_{mn}(\bk)}
  u_{n,\bk} u_{m,\bk}^*\langle
  u_{n,\bk}|i\nabla_{\bk}|u_{m,\bk} \rangle_{L^2(\Gamma)}\rd\bk \\
  &+\sum_{n\leq Z}\sum_{m>Z}\barint_{\Gamma^*}\frac{1}{\omega -
    \omega_{mn}(\bk)} u_{n,\bk}^* u_{m,\bk}\overline{\langle
    u_{n,\bk}|i\nabla_{\bk}|u_{m,\bk} \rangle}_{L^2(\Gamma)}\rd\bk.
\end{align*}
Remark that the dynamic permittivity matrix $\mc{E}$ is completely
determined by the linear response of the unperturbed electronic
structure.

The Maxwell equations \eqref{eqn:E}-\eqref{eqn:B} are nonlocal in
time, since the permittivity $\mc{E}$ depends on $\omega$. While the
system is local in space due to the limit $\veps \to 0$, the
nonlocality in time is natural since there is no scale separation in
time.

We also note that while we obtain a nontrivial effective
permittivity, the effective permeability in the equation equals to
$1$, the same value as in the vacuum. Physically, this is consistent
with the situation of semiconductors or insulators under
consideration here.

\section{Asymptotic analysis of the Schr\"odinger-Maxwell equations}
\label{sec:asymptotics}

To derive the effective Maxwell equation, let us take the following
ansatz for the system \eqref{eqn:Schr}-\eqref{eqn:V_tot},
\begin{align}\label{ansatz:rho}
  & \rho^{\veps}(t,\bx)=\veps^{-3}\rho_0(\bx/\veps)+\veps^{-2}
  \rho_1(t,\bx,\bx/\veps)+\veps^{-1}\rho_2(t,\bx,\bx/\veps)+\cdots\;,\\
  \label{ansatz:J}
  & \bJ^{\veps}(t,\bx)=\veps^{-3}\bJ_{0}(\bx/\veps)+\veps^{-2}
  \bJ_1(t,\bx,\bx/\veps)+\cdots\; , \\
  \label{ansatz:A} &
  \bA^\veps(t,\bx)+\bA_{\ext}(t,\bx)=\bA_0(t,\bx,\bx/\veps)
  +\veps\bA_1(t,\bx,\bx/\veps) +\cdots\; , \\
  \label{ansatz:phi}
  &\phi^\veps(t,\bx)+V_{\ext}(t,\bx)=\phi_{0}(t,\bx,\bx/\veps)+\veps
  \phi_{1}(t,\bx,\bx/\veps) \\
  &\hspace{18em}+\veps^{2} \phi_{2}(t,\bx,\bx/\veps)+\cdots\; ,\nn
\end{align}
where the higher order terms are omitted. We also assume that the
dependence on the fast variable $\bz = \bx/\veps$ is periodic for all
these functions.

The main strategy of asymptotic analysis is as follows. We first
apply a two-scale expansion on the Maxwell equations
\eqref{eqn:Poisson}-\eqref{eqn:VecPot}, which produces the asymptotics
of Hamiltonian; then by Dyson series we obtain the asymptotics of
density and current; the effective equations in time domain are
derived by taking the $\bz$-average on the second order perturbation
equations of the Coulomb potential and vector potential.  The
asymptotics is somewhat nontrivial. The Coulomb interaction makes the
leading order potential dependent on the macroscopic average of the third
order density. To close the asymptotics, one has to show that the
macroscopic average of the third order density only depends on the
leading order potential, but not on higher order terms of the
potential. This amounts to establishing the local neutrality of the system, which
will be explained in detail below.  Finally Fourier transform gives
the effective Maxwell equation in frequency domain. Note that we have
assumed that the leading order density and current only depend on the
fast variable $\bx/\veps$. This will be justified by the asymptotics.

\subsection{Asymptotics of the Hamiltonian}

For the Coulomb potential, substituting the ansatz
\eqref{ansatz:rho} and \eqref{ansatz:phi} in \eqref{eqn:Poisson} and
organizing the results in orders, one has
\begin{align}
  & -\Delta_{\bz}\phi_0 = \rho_0-m_0,\label{asym:phi0}\\
  & -\Delta_{\bz}\phi_1-2\nabla_{\bx}\cdot\nabla_{\bz}\phi_0
  =\rho_1, \label{asym:phi1} \\
  & -\Delta_{\bz}\phi_2-2\nabla_{\bx}\cdot\nabla_{\bz}\phi_{1}
  -\Delta_{\bx}\phi_0 =\rho_2+\rho_{\ext}.\label{asym:phi2}
\end{align}
Recall that $\rho_{\ext}(t,\bx)=-\Delta_{\bx}V_{\ext}(t,\bx)$.

For the exchange-correlation potential, Taylor expansion yields
\begin{equation}
  \begin{aligned}
    \eta(\veps^3\rho^{\veps}) &= \eta(\rho_0)+\veps
    \eta'(\rho_0)\rho_1+\tfrac{1}{2}\veps^2\eta''(\rho_0)\rho_1^2 +
    \veps^2 \eta'(\rho_0)\rho_2 +\cdots \\
    & =\eta_0+\veps\eta_1+\veps^2\eta_2+\cdots\;,
  \end{aligned}
\end{equation}
where the last equality gives the definition of $\eta_i(t, \bx,
\bz)$,
\begin{align*}
  & \eta_0(\bz) = \eta(\rho_0(\bz)), \\
  & \eta_1(t, \bx, \bz) = \eta'(\rho_0(\bz))\rho_1(t, \bx, \bz), \\
  & \eta_2(t, \bx, \bz) = \tfrac{1}{2} \eta''(\rho_0(\bz))\rho_1(t,
  \bx, \bz)^2 + \eta'(\rho_0(\bz))\rho_2(t, \bx, \bz),
\end{align*}
and similarly for higher order terms, which we omitted in the
expression.

Therefore, the total potential $V^\veps$ can be written as
\begin{equation}\label{ansatz:V}
\begin{aligned}
  V^\veps &=\phi+V_{\ext}+\eta\\
  & =(\phi_0+\eta_0)+\veps(\phi_1+\eta_1)
  +\veps^2(\phi_2+\eta_2)+\cdots  \\
  &=V_0+\veps V_1 +\veps^2 V_2 + \cdots \; .
\end{aligned}
\end{equation}

Similarly, we write down the equations for the vector potential using
ansatz \eqref{ansatz:J} and \eqref{ansatz:A}:
\begin{align}
  & -\Delta_{\bz} \bA_0 = 0, \label{asym:A0}\\
  & -\Delta_{\bz}\bA_1 - 2\nabla_{\bx}\cdot\nabla_{\bz}\bA_0
  +\frac{\partial}{\partial t}\bigl(\nabla_{\bz}
  V_0\bigr) = \bJ_{0}, \label{asym:A1}\\
  &   \label{asym:A2} \frac{\partial^2}{\partial
    t^2}\bA_0-\Delta_{\bx}\bA_0-\nabla_{\bz}\bA_2
  -2\nabla_{\bx}\cdot\nabla_{\bz}\bA_1 \\
  & \hspace{6em} +\frac{\partial}{\partial t}\bigl(\nabla_{\bz}
  V_1+\nabla_{\bx} V_0\bigr) = \bJ_1+\bJ_{\ext}. \nn
\end{align}
Recall that $\bJ_{\ext}(t,\bx)=\frac{\partial^2}{\partial
  t^2}\bA_{\ext} -\Delta_{\bx} \bA_{\ext}+ \frac{\partial}{\partial
  t}\left(\nabla_{\bx} V_{\ext}\right)$.

Note that the solvability conditions of
\eqref{asym:phi0}-\eqref{asym:phi1} and \eqref{asym:A1} impose the
following constraint on $\rho_0$, $\rho_1$ and $\bJ_{0}$,
\begin{equation}\label{aspt:rho}
  \langle\rho_0(\bz)\rangle_{\bz}=\langle
  m_0(\bz)\rangle_{\bz},\quad
  \langle\rho_1(t,\bx,\bz)\rangle_{\bz}=0,\quad
  \average{\bJ_{0}(\bz)}_{\bz}=0.
\end{equation}

Therefore, the Hamiltonian operator can be written as
\begin{equation}\label{op:Hamil}
  \begin{aligned}
    H^\veps(t) = -\tfrac{1}{2} \veps^2 \Delta_{\bx} & + v_0(\bx/\veps)
    + U_0(t,\bx) + \veps v_1(t,\bx,\bx/\veps) \\
    & + \veps U_1(t,\bx) +i\veps^2 \bA_0(t,\bx)\cdot\nabla_{\bx}
    +\veps^2 V_2(t,\bx,\bx/\veps) \\
    & +\veps^2\frac{\abs{\bA_0(t,\bx)}^2}{2} +
    i\veps^3\bA_1(t,\bx,\bx/\veps)\cdot\nabla_{\bx} + \cdots
  \end{aligned}
\end{equation}
where we omit the higher order terms. In \eqref{op:Hamil}, we define
$v_0,\;v_1$ and $U_0,\;U_1$ as the {\it microscopic} and {\it
  macroscopic} components of $V_0,\;V_1$ respectively,
\begin{equation}\label{eqn:potential_decomp}
  \begin{aligned}
    & U_0(t, \bx) = \average{V_0(t,\bx,\cdot)}_{\bz},
    && v_0(t, \bx, \bz) = V_0(t, \bx, \bz) - U_0(t, \bx); \\
    & U_1(t, \bx) = \average{V_1(t,\bx,\cdot)}_{\bz}, && v_1(t, \bx,
    \bz) = V_1(t, \bx, \bz) - U_1(t, \bx).
  \end{aligned}
\end{equation}

\subsection{Asymptotics of the density and current}

The initial state is given by the ground state of the unperturbed
system, which implies that the density matrix is given by the
projection operator to the occupied spectrum of the ground states in
the beginning (with lattice parameter $\veps$),
\begin{equation}
  \bvec{\rho}^{\veps}(0)=\mc{P}^{\veps}.
\end{equation}
Then, the density matrix at time $t$ is given by\footnote{In the
language of physics, we are using the Heisenberg picture.}
\begin{equation}\label{op:DenMat}
  \bvec{\rho}^{\veps}(t) = \mathcal{T} \exp\biggl(-i\int_0^t H^\veps(\tau)\rd\tau
  \biggr)\mathcal{P}^{\veps}
  \biggl(\mathcal{T} \exp\biggl(-i\int_0^t H^\veps(\tau)\rd \tau\biggr)
  \biggr)^{\ast},
\end{equation}
where $\mathcal{T}$ is the time ordering operator. Therefore the
density is given by the diagonal of the kernel of the operator
$\bvec{\rho}^{\veps}$,
\begin{multline}\label{eqn:density}
  \rho^{\veps}(t,\bx) = \bvec{\rho}^{\veps}(t,\bx,\bx) = \mathcal{T}
  \exp\biggl(-i\int_0^t H^\veps(\tau)\rd\tau\biggr) \\
  \times \mathcal{P}^{\veps}
  \biggl(\mathcal{T} \exp\biggl(-i\int_0^t H^\veps(\tau)\rd\tau\biggr)
  \biggr)^{\ast}(\bx,\bx),
\end{multline}
where the right hand side means the diagonal of the kernel associated
with the operator. The ground state density is given by
$\rho_{\gs}^{\veps}(\bx)=\mc{P}^{\veps}(\bx,\bx)$.

We investigate the asymptotic expansion of $\rho^{\veps}(t,\bx)$
determined by the Hamiltonian \eqref{op:Hamil}. The equation
\eqref{eqn:density} implies, to obtain the density at $(t,\bx)$, we
first evolve the system backwards to the initial time, project it
onto the ground states of the unperturbed system, then evolve it
forwards in time to $t$ at $\bx$. Since the current time scaling is
$\Or(1)$, when $\veps$ goes to $0$, the domain of dependence and
domain of influence are also of the scale $\Or(\veps)$ for the
system in the time evolution. In other words, the density at
$(t,\bx)$ only depends on the Hamiltonian of a small neighborhood
$(0,t)\times B(\bx, \Or(\veps))$, where $B(\bx, \Or(\veps))$
indicates a ball centered at $\bx$ with the radius $\Or(\veps)$.

%

Accordingly for a fixed point $\bx \in \Gamma$, we expand the
Hamiltonian operator $H$ around $\bx$. For clarity, we write $H$ as
an operator on $L_{\by}^2(\Gamma)$,
\begin{equation*}
  \begin{aligned}
    H^\veps(t) = -\tfrac{1}{2} \veps^2 \Delta_{\by} & + v_0(\by/\veps)
    + U_0(t,\by) + \veps v_1(t,\by,\by/\veps) \\
    & + \veps U_1(t,\by) +i\veps^2 \bA_0(t,\by)\cdot\nabla_{\by}
    +\veps^2 V_2(t,\by,\by/\veps) \\
    & +\veps^2\frac{\abs{\bA_0(t,\by)}^2}{2} +
    i\veps^3\bA_1(t,\by,\by/\veps)\cdot\nabla_{\by} + \cdots.
  \end{aligned}
\end{equation*}
Let $H_0^{\veps}(t, \bx)$ be the leading order operator
\begin{equation}\label{op:H0}
  H_0^{\veps}(t,\bx) = - \tfrac{1}{2} \veps^2 \Delta_{\by} +
  v_0(\by/\veps) + U_0(t,\bx).
\end{equation}
Here $H_0^{\veps}(t, \bx)$ is an operator on $L_{\by}^2$ with $(t,
\bx)$ as parameters. Similar notations will be used throughout the
paper. Denote the difference as $\delta H^{\veps}(t,\bx) =
H^{\veps}(t) - H_0^{\veps}(t, \bx)$,
\begin{equation}
  \begin{aligned}
    \delta H^{\veps}(t,\bx) & = U_0(t,\by) - U_0(t,\bx) + \veps
    v_1(t,\by, \by/\veps) + \veps U_1(t,\by) \\
    & +i\veps^2 \bA_0(t,\by)\cdot\nabla_{\by} + \veps^2
    V_2(t,\by,\by/\veps)
    +\veps^2\frac{\abs{\bA_0(t,\by)}^2}{2} \\
    & +i\veps^3\bA_1(t,\by,\by/\veps)\cdot\nabla_{\by} + \cdots.
  \end{aligned}
\end{equation}
Expand $\delta H^{\veps}(t,\bx)$ into orders, and the first two
orders are
\begin{equation}\label{op:H1}
  \delta H_1^{\veps}(t,\bx) = (\by-\bx)\cdot\nabla_{\bx} U_0(t,\bx)
  + \veps v_1(t,\bx,\by/\veps) + \veps U_1(t,\bx)+i\veps \bA_0(t,\bx)
  \cdot\veps\nabla_{\by},
\end{equation}
and
\begin{equation}\label{op:H2}
  \begin{aligned}
    \delta H_2^{\veps}(t,\bx) =&
    \tfrac{1}{2}\big((\by-\bx)\cdot\nabla_{\bx}\big)^2 U_0(t,\bx)
    + \veps (\by-\bx)\cdot\nabla_{\bx} v_1(t,\bx,\by/\veps) \\
    & + \veps (\by-\bx)\cdot\nabla_{\bx} U_1(t,\bx) + \veps^2
    V_2(t,\bx,\by/\veps)+\veps^2\frac{\abs{\bA_0(t,\bx)}^2}{2}\\
    & + i\bigl(\veps(\by-\bx)\cdot\nabla_{\bx}\bA_0(t,\bx)
    +\veps^2\bA_1(t,\bx,\by/\veps)\bigr)\cdot\veps\nabla_{\by}.
  \end{aligned}
\end{equation}
Recall that we are approximating the Hamiltonian in a ball around
$\bx$ with the radius $\Or(\veps)$, and hence $(\by - \bx)$ is treated
as $\Or(\veps)$ in the above expansion.

We expand the time propagation operator as
\begin{align*}
  \mathcal{T} \exp\biggl(-i\int_0^t H^{\veps}(\tau)\rd\tau\biggr) =&
   \mc{U}_{t,0}^{\veps,0}(\bx) \\
  & - i \int_0^t \mathcal{T} \exp\biggl(-i\int_{\tau}^t
  H^{\veps}(s)\rd s\biggr)
  \delta H^{\veps}(\tau,\bx) \mathcal{U}_{\tau,0}^{\veps,0}(\bx) \rd \tau \\
  = & \mc{U}_{t,0}^{\veps,0}(\bx) - i \int_0^t
  \mc{U}_{t,\tau}^{\veps,0}(\bx)
  \delta H^{\veps}(\tau,\bx) \mathcal{U}_{\tau,0}^{\veps,0}(\bx) \rd \tau \\
  & - \int_0^t \int_0^{\tau_2} \mc{U}_{t,\tau_2}^{\veps,0}(\bx) \delta
  H^{\veps}(\tau_2,\bx)
  \mc{U}_{\tau_2,\tau_1}^{\veps,0}(\bx) \\
  & \qquad \times \delta H^{\veps}(\tau_1,\bx)
  \mc{U}_{\tau_1,0}^{\veps,0}(\bx) \rd\tau_1 \rd\tau_2 \\
  & + \cdots,
\end{align*}
where $\mathcal{U}_{t_1,t_2}^{\veps,0}(\bx)$ is the propagation operator
corresponding to $H_0^{\veps}$,
\begin{equation*}
  \mathcal{U}_{t_1,t_2}^{\veps,0}(\bx) = \mathcal{T} \exp\biggl(-i\int_{t_2}^{t_1}
  H_0^\veps(\tau,\bx) \rd\tau\biggr).
\end{equation*}
Therefore by \eqref{op:H1}-\eqref{op:H2},
\begin{equation}\label{expansion:PropOp}
  \begin{aligned}
    \mathcal{T} \exp\biggl(-i\int_0^t H^{\veps}(\tau)\biggr)\rd\tau =&
    \mc{U}_{t,0}^{\veps,0}(\bx) - i \int_0^t
    \mc{U}_{t,\tau}^{\veps,0}(\bx)
    \delta H_1^{\veps}(\tau,\bx) \mathcal{U}_{\tau,0}^{\veps,0}(\bx) \rd \tau \\
    & - i \int_0^t \mc{U}_{t,\tau}^{\veps,0}(\bx)
    \delta H_2^{\veps}(\tau,\bx) \mathcal{U}_{\tau,0}^{\veps,0}(\bx) \rd \tau \\
    & - \int_0^t \int_0^{\tau_2} \mc{U}_{t,\tau_2}^{\veps,0}(\bx)
    \delta H_1^{\veps}(\tau_2,\bx) \mc{U}_{\tau_2,\tau_1}^{\veps,0}(\bx) \\
    & \qquad \times \delta H_1^{\veps}(\tau_1,\bx)
    \mc{U}_{\tau_1,0}^{\veps,0}(\bx) \rd\tau_1 \rd\tau_2 + \cdots \\
    = & \mc{U}_{t,0}^{\veps,0}(\bx) + \veps
    \mc{U}_{t,0}^{\veps,1}(\bx) + \veps^2 \mc{U}_{t,0}^{\veps,2}(\bx)
    + \cdots,
  \end{aligned}
\end{equation}
where the last equality defines $\mc{U}_{t,0}^{\veps,j}(\bx)$ for $j =
0, 1, 2$. Higher order terms can be written down similarly.

Substituting the expansion \eqref{expansion:PropOp} in
\eqref{eqn:density}, we obtain to the leading order
\begin{equation*}
  \rho_0^{\veps}(t,\bx) =
  \Bigl(\mc{U}^{\veps,0}_{t,0}(\bx) \mathcal{P}^{\veps}
  \mc{U}^{\veps,0}_{0,t}(\bx)\Bigr)(\bx,\bx),
\end{equation*}
and also to the higher orders,
\begin{align*}
  \rho_1^{\veps}(t,\bx) & = \Bigl(\mc{U}^{\veps,0}_{t,0}(\bx)
  \mathcal{P}^{\veps} \mc{U}^{1,\veps}_{0,t}(\bx)\Bigr)(\bx,\bx) +
  \Bigl(\mc{U}^{\veps,1}_{t,0}(\bx) \mathcal{P}^{\veps}
  \mc{U}^{\veps,0}_{0,t}(\bx)\Bigr)(\bx,\bx), \\
  \rho_2^{\veps}(t,\bx) & = \Bigl(\mc{U}^{\veps,0}_{t,0}(\bx)
  \mathcal{P}^{\veps} \mc{U}^{2,\veps}_{0,t}(\bx)\Bigr)(\bx,\bx) +
  \Bigl(\mc{U}^{\veps,1}_{t,0}(\bx) \mathcal{P}^{\veps}
  \mc{U}^{\veps,1}_{0,t}(\bx)\Bigr)(\bx,\bx) \\
  & + \Bigl(\mc{U}^{\veps,2}_{t,0}(\bx) \mathcal{P}^{\veps}
  \mc{U}^{\veps,0}_{0,t}(\bx)\Bigr)(\bx,\bx) \nn.
\end{align*}
Therefore the density is given by
\begin{equation*}
  \rho^{\veps}(t,\bx) = \rho_0^{\veps}(t,\bx) +
  \veps \rho_1^{\veps}(t,\bx) + \veps^2 \rho_2^{\veps}(t,\bx) + \Or(\veps^3).
\end{equation*}

The current is given by
\begin{equation*}
\bJ^{\veps}(t,\bx)=\frac{1}{2i}\big[\veps\nabla_{\by},\;
\bvec{\rho}^{\veps}(t)\big]_{+}(\bx,\bx),
\end{equation*}
where $\bvec{\rho}^\veps(t)$ is the density matrix in
\eqref{op:DenMat} and $[a, \; b]_{+}=ab+ba$ is the anticommutator.

With the help of \eqref{expansion:PropOp}, similarly one has
\[\bJ^{\veps}(t,\bx)=\bJ_0^{\veps}(t,\bx)+\veps\bJ_1^{\veps}(t,\bx)+\Or(\veps^2),\]
where
\begin{align*}
& \bJ_0^\veps=\frac{1}{2i}\bigg[ \veps\nabla_{\by},\;
\mc{U}^{\veps,0}_{t,0}(\bx) \mathcal{P}^{\veps}
  \mc{U}^{\veps,0}_{0,t}(\bx)
\bigg]_{+}(\bx,\bx),\\ & \bJ_1^\veps=\frac{1}{2i}\bigg[
\veps\nabla_{\by},\; \mc{U}^{\veps,0}_{t,0}(\bx)
  \mathcal{P}^{\veps} \mc{U}^{1,\veps}_{0,t}(\bx)+
  \mc{U}^{\veps,1}_{t,0}(\bx) \mathcal{P}^{\veps}
  \mc{U}^{\veps,0}_{0,t}(\bx)
\bigg]_{+}(\bx,\bx)-\rho_0^{\veps}\bA_0(t, \bx).
\end{align*}

\subsection{Rescalings of Hamiltonian, density and current}
Notice that the operator $H_0^{\veps}(t,\bx)$ in \eqref{op:H0} can
be rescaled as
\begin{equation*}
  \delta_{\veps} H_0^{\veps}(t,\bx) \delta^{\ast}_{\veps} = H_0(t,\bx),
\end{equation*}
with $\delta_{\veps}$ as the dilation operator and $H_0$ is given by
\begin{equation*}
  H_0(t,\bx) = - \tfrac{1}{2}\Delta_{\bzeta} + v_0(\bzeta) +
  U_0(t,\bx),
\end{equation*}
where $\bzeta=\by/\veps$ is the small scale spatial variable.

Therefore, we can rescale the expressions of
$\rho^{\veps}_{j}(t,\bx)$ and $\bJ^{\veps}_{j}(t,\bx)$ by
\begin{equation*}
  \rho^{\veps}_{j}(t,\bx) = \veps^{-3}\rho_{j}(t,\bx,\bx/\veps), \quad
\bJ^{\veps}_{j}(t,\bx)=\veps^{-3}\bJ_{j}(t,\bx,\bx/\veps), \quad j =
0, 1, \cdots,
\end{equation*}
and
\begin{align}
  \rho_0(t,\bx,\bz) = & \; \Bigl(\mc{U}^0_{t,0}(\bx) \mc{P}
  \mc{U}^0_{0,t}(\bx)\Bigr)(\bz,\bz), \\
  \rho_1(t,\bx,\bz) = & \; \Bigl(\mc{U}^0_{t,0}(\bx) \mc{P}
  \mc{U}^1_{0,t}(\bx,\bz)\Bigr)(\bz,\bz) + \Bigl(\mc{U}^1_{t,0}(\bx,\bz) \mc{P}
  \mc{U}^0_{0,t}(\bx)\Bigr)(\bz,\bz),
  \label{eqn:rho1_old}\\
  \rho_2(t,\bx,\bz) = & \; \Bigl(\mc{U}^0_{t,0}(\bx) \mc{P}
  \mc{U}^2_{0,t}(\bx,\bz)\Bigr)(\bz,\bz) + \Bigl(\mc{U}^1_{t,0}(\bx,\bz) \mc{P}
  \mc{U}^1_{0,t}(\bx,\bz)\Bigr)(\bz,\bz)
  \label{eqn:rho2_old}\\
  & + \Bigl(\mc{U}^2_{t,0}(\bx,\bz) \mc{P} \mc{U}^0_{0,t}(\bx)\Bigr)(\bz,\bz)
  \nn, \\ \label{eqn:J0_old}
  \bJ_0(t,\bx,\bz)=&\frac{1}{2i}\bigg[\nabla_{\bzeta},\;
  \mc{U}^{0}_{t,0}(\bx) \mathcal{P}
  \mc{U}^{0}_{0,t}(\bx)\bigg](\bz,\bz), \\ \label{eqn:J1_old}
  \bJ_1(t,\bx,\bz)=&\frac{1}{2i}\bigg[\nabla_{\bzeta},\;
  \mc{U}^{0}_{t,0}(\bx)
  \mathcal{P} \mc{U}^{1}_{0,t}(\bx,\bz)+
  \mc{U}^{1}_{t,0}(\bx,\bz) \mathcal{P}
  \mc{U}^{0}_{0,t}(\bx)\bigg](\bz,\bz)\\&
  -\rho_0(t,\bx,\bz)\bA_0(t,\bx),\nn
\end{align}
where $\mc{P}$ is the rescaled (lattice parameter $1$) density
matrix for the unperturbed system, and $\mc{U}^i_{t,s}$ are
propagation operators defined on $L^2_{\bzeta}$ by
\begin{equation*}
  \mc{U}^0_{t,s}(\bx) = \mc{T} \exp(-i\int_s^t H_0(\tau,\bx)
  \rd\tau),
\end{equation*}
\begin{equation}\label{op:U1}
  \mc{U}^1_{t,s}(\bx,\bz) = -i \int_s^t \mc{U}^0_{t,\tau}(\bx)
  \delta H_1(\tau, \bx, \bz) \mc{U}^0_{\tau, 0}(\bx) \rd \tau,
\end{equation}
\begin{equation}\label{op:U2}
  \begin{aligned}
    \mc{U}^2_{t,s}(\bx, \bz) & = - i \int_s^t \mc{U}^0_{t,\tau}(\bx)
    \delta H_2(\tau, \bx, \bz)
    \mc{U}^0_{\tau, 0}(\bx) \rd \tau \\
    & - \int_s^t\int_s^{\tau_2} \mc{U}^0_{t,\tau_2}(\bx) \delta
    H_1(\tau_2,\bx, \bz) \mc{U}^0_{\tau_2,\tau_1}(\bx) \delta
    H_1(\tau_1,\bx, \bz) \mc{U}^0_{\tau_1,0}(\bx) \rd \tau_1 \rd
    \tau_2,
  \end{aligned}
\end{equation}
in which
\begin{align}
  H_0(t,\bx) & = -\tfrac{1}{2}\Delta_{\bzeta}+ v_0(\bzeta) + U_0(t,\bx), \\
  \label{op:dH_1} \delta H_1(t,\bx,\bz) & = (\bzeta-\bz)\cdot\nabla_{\bx}
  U_0(t,\bx)+v_1(t,\bx,\bzeta)+U_1(t,\bx)+i\bA_0\cdot\nabla_{\bzeta}, \\
  \label{op:dH_2}\delta H_2(t,\bx,\bz) & =
  \tfrac{1}{2}\big((\bzeta-\bz)\cdot \nabla_{\bx}\big)^2 U_0(t,\bx)\\
  &\quad+(\bzeta-\bz)\cdot\nabla_{\bx}
  \big(v_1(t,\bx,\bzeta)+U_1(t,\bx)\big)+V_2(t,\bx,\bzeta)\nn \\
  &\quad+i\bigg((\bzeta-\bz)\cdot\nabla_{\bx}\bA_0+\bA_1\bigg)\cdot\nabla_{\bzeta}
  +\tfrac{1}{2}\abs{\bA_0}^2.\nn
\end{align}


Notice that the dependence of the operator $H_0(\tau,\bx)$ on $\bx$
lies only in $U(\tau,\bx)$ which works like a number as an operator
on $L^2_{\bzeta}$. It implies
\begin{equation*}
  \mc{U}^0_{t,s}(\bx) = \exp\biggl(-i\int_s^t U(\tau,\bx)\biggr)
  \exp(-i (t-s) H_0),
\end{equation*}
where $H_0$ agrees with the unperturbed Hamiltonian
\begin{equation}\label{op:H0_z}
    H_0 = -\tfrac{1}{2}\Delta_{\bzeta} + v_0(\bzeta).
\end{equation}
Moreover, in the expression of $\rho_j$, the phase factor
$\exp(-i\int_s^t U(\tau,\bx))$ will not appear since it gets
canceled by its complex conjugate. Therefore, we can simply take
\begin{equation}\label{op:U0}
    \mc{U}^0_{t,s} = \exp(-i(t-s) H_0),
\end{equation}
which is, in particular, independent of $\bx$.

It results that the leading order density agrees with the ground
state electron density of the unperturbed system,
\begin{equation}\label{eqn:rho0_avg}
    \rho_0(t,\bx,\bz)= e^{-it H_0} \mathcal{P} e^{it H_0}(\bz,\bz) = \rho_{\gs}(\bz).
\end{equation}
This also shows the following proposition.

\begin{prop}
  $\rho_0$ is {\it independent} of $t$ and $\bx$ and
  \[\average{\rho_0}_{\bz}=\average{m_0}_{\bz},\]
  which satisfies the first constraint of \eqref{aspt:rho}
  self-consistently.
\end{prop}

\section{Effective equations in the time domain}\label{sec:time}

In this section, we derive the main result in time domain. It will
be connected to the main result described in frequency domain in the
next Section.

\subsection{Effective equations in time domain}

Let us first summarize the resulting equations in the time domain.
Recall that
\begin{align*}
  & V_0(t,\bx,\bz)=v_0(\bz)+U_0(t,\bx), \\
  & V_1(t,\bx,\bz)=v_1(t,\bx,\bz)+U_1(t,\bx), \\
  & \bA_0(t,\bx,\bz)=\bA_0(t,\bx),
\end{align*}
then one has the following effective equations from
\eqref{eqn:Schr}-\eqref{eqn:V_tot}.
\begin{itemize}
\item The microscopic scalar potential $v_0(\bz)$ is the same as that
  of the unperturbed system.

\item The potentials $v_1(t, \bx, \bz)$, $U_0(t, \bx)$ and $\bA_0(t,
  \bx)$ form a closed system as described below. The microscopic
  scalar potential $v_1(t,\bx,\bz)$ is given by
  \begin{equation}\label{eqn:V1}
    \begin{aligned}
      & v_1=V_1(t,\bx,\bz)-\average{V_1(t,\bx,\bz)}_{\bz}, \\
      & V_1=\mathcal{V}\rho_1=\phi_1+\eta'(\rho_0)\rho_1, \\
      & -\Delta_{\bz}\phi_1 = \rho_1.
    \end{aligned}
  \end{equation}
  Recall that $\mathcal{V}$ is the linearization of the effective
  potential operator at equilibrium density. Here, the density
  $\rho_1$ is given by the equation
  \begin{equation}\label{eqn:rho1_thm}
    (I-\chi\mathcal{V})\rho_1=
    \int_0^t\bbf(t-\tau)\cdot\nabla_{\bx}U_0(\tau)\rd\tau+\int_0^t
    \bbg(t-\tau)\cdot\bA_0(\tau)\rd\tau,
  \end{equation}
  where $I$ is the identity operator,
  \begin{align*}
    &\chi\mathcal{V} \rho_1 =\chi v_1 =-2\Im \sum_{n\leq Z}\sum_{m>Z}
    \int_0^t \barint_{\Gamma^*} e^{i\omega_{mn}(\bk)(t-\tau)} \\&
    \hspace{11em} \times u_{n,\bk} u^{\ast}_{m,\bk}\langle
    u_{n,\bk}|v_1(\tau)|u_{m,\bk} \rangle_{L^2(\Gamma)}\rd\bk \rd\tau,
  \end{align*}
  and
  \begin{align*}
    & \bbf(s)=-2\Im \sum_{n\leq Z}\sum_{m>Z} \barint_{\Gamma^*}
    e^{i\omega_{mn}(\bk)s} u_{n,\bk}u^{\ast}_{m,\bk} \langle
    u_{n,\bk}|i\nabla_{\bk}|u_{m,\bk}
    \rangle_{L^2(\Gamma)}\rd\bk, \\
    & \bbg(s)=-2\Im \sum_{n\leq Z}\sum_{m>Z} \barint_{\Gamma^*}
    e^{i\omega_{mn}(\bk)s} u_{n,\bk}u^{\ast}_{m,\bk} \langle
    u_{n,\bk}|i\nabla_{\bzeta}|u_{m,\bk}
    \rangle_{L^2(\Gamma)}\rd\bk.
  \end{align*}
  Here we have introduced the short hand notation
  \begin{equation*}
    \omega_{mn}(\bk) = E_m(\bk) - E_n(\bk).
  \end{equation*}

\item The macroscopic scalar potential $U_0(t,\bx)$ satisfies
  \begin{equation}\label{eqn:U0}
    - \Delta U_0 - P_{\alpha\beta}  \partial_{\bx_\alpha}\partial_{\bx_\beta}U_0 =
    Q_\alpha \left(\partial_{\bx_\alpha} v_1\right)
    + R_{\alpha\beta}\bigl(\partial_{\bx_\alpha}(\bA_0)_{\beta}\bigr)
    + \rho_{\ext}(t,\bx).
  \end{equation}
  Here
  \begin{multline*}
    \qquad
    P_{\alpha\beta}\left(\partial_{\bx_\alpha}\partial_{\bx_\beta}U_0\right)=
    2\Im \sum_{n\leq Z}\sum_{m>Z}\int_0^t\barint_{\Gamma^*}
    e^{i\omega_{mn}(\bk)(t-\tau)} \overline{\langle
      u_{n,\bk}|i\partial_{\bk_\alpha}|u_{m,\bk}
      \rangle}_{L^2(\Gamma)}\\ \times \langle
    u_{n,\bk}|i\partial_{\bk_\beta}|u_{m,\bk}
    \rangle_{L^2(\Gamma)}\partial_{\bx_\alpha}\partial_{\bx_\beta}U_0(\tau)\rd\bk\rd\tau
    ,
  \end{multline*}
  \begin{multline*}
    \qquad Q_\alpha \left(\partial_{\bx_\alpha} v_1\right)= 2\Im
    \sum_{n\leq Z} \sum_{m>Z} \int_0^t\barint_{\Gamma^*}
    e^{i\omega_{mn}(\bk)(t-\tau)} \overline{\langle
      u_{n,\bk}|i\partial_{\bk_\alpha}|u_{m,\bk}
      \rangle}_{L^2(\Gamma)} \\ \times \langle
    u_{n,\bk}|\partial_{\bx_\alpha} v_1(\tau)|u_{m,\bk}
    \rangle_{L^2(\Gamma)}\rd\bk\rd\tau,
  \end{multline*}
  and
  \begin{multline*}
    \qquad R_{\alpha\beta} \bigl(\partial_{\bx_\alpha}
    (\bA_0)_{\beta}\bigr)= 2\Im \sum_{n\leq Z} \sum_{m>Z}
    \int_0^t\barint_{\Gamma^*} e^{i\omega_{mn}(\bk)(t-\tau)}
    \overline{\langle u_{n,\bk}|i\partial_{\bk_\alpha}|u_{m,\bk}
      \rangle}_{L^2(\Gamma)} \\ \times {\langle
      u_{n,\bk}|i\partial_{\bzeta_{\beta}}|u_{m,\bk}}
    \rangle_{L^2(\Gamma)}\partial_{\bx_\alpha}(\bA_0)_{\beta}\rd\bk\rd\tau.
  \end{multline*}

\item The macroscopic vector potential $\bA_0$ satisfies
  \begin{align}\label{eqn:A0}
    &\frac{\partial^2}{\partial
      t^2}(\bA_0)_{\alpha}-\Delta_{\bx}(\bA_0)_{\alpha}+\frac{\partial}{\partial
      t}(\partial_{\bx_\alpha}U_0) = \\
    &\qquad
    S_\alpha(v_1)+M_{\alpha\beta}(\partial_{\bx_\beta}U_0)+N_{\alpha\beta}
    \bigl((\bA_0)_{\beta}\bigr)
    -\average{\rho_0}_z(\bA_0)_{\alpha}+(\bJ_{\ext})_{\alpha}(t,\bx), \nn\\
    &\nabla_{\bx}\cdot \bA_0=0.\nn
  \end{align}
  Here
  \begin{multline*}
    \qquad S_\alpha (v_1)= 2\Im \sum_{n\leq Z} \sum_{m>Z}
    \int_0^t\barint_{\Gamma^*} e^{i\omega_{mn}(\bk)(t-\tau)}
    \overline{\langle u_{n,\bk}|i\partial_{\bzeta_\alpha}|u_{m,\bk}
      \rangle}_{L^2(\Gamma)} \\ \times {\langle u_{n,\bk}|
      v_1(\tau)|u_{m,\bk}} \rangle_{L^2(\Gamma)}\rd\bk\rd\tau,
  \end{multline*}
  \begin{multline*}
    \qquad M_{\alpha\beta}\left(\partial_{\bx_\beta}U_0\right)= 2\Im
    \sum_{n\leq Z}\sum_{m>Z}\int_0^t\barint_{\Gamma^*}
    e^{i\omega_{mn}(\bk)(t-\tau)} \overline{\langle
      u_{n,\bk}|i\partial_{\bzeta_\alpha}|u_{m,\bk}
      \rangle}_{L^2(\Gamma)} \\ \times \langle
    u_{n,\bk}|i\partial_{\bk_\beta}|u_{m,\bk}
    \rangle_{L^2(\Gamma)}\partial_{\bx_\beta}U_0(\tau)\rd\bk\rd\tau ,
  \end{multline*}
  and
  \begin{multline*}
    \qquad N_{\alpha\beta}\bigl((\bA_0)_\beta\big)= 2\Im \sum_{n\leq
      Z}\sum_{m>Z}\int_0^t\barint_{\Gamma^*}
    e^{i\omega_{mn}(\bk)(t-\tau)} \overline{\langle
      u_{n,\bk}|i\partial_{\bzeta_\alpha}|u_{m,\bk}
      \rangle}_{L^2(\Gamma)} \\ \times \langle
    u_{n,\bk}|i\partial_{\bzeta_\beta}|u_{m,\bk}
    \rangle_{L^2(\Gamma)}(\bA_0)_\beta(\tau)\rd\bk\rd\tau.
  \end{multline*}
\end{itemize}
The above equations from \eqref{eqn:V1} to \eqref{eqn:A0} form a
{\it closed} system that determines the macroscopic potentials
$U_0(t, \bx)$ and $\bA_0(t, \bx)$.

\subsection{Derivation}
We first give a description on how to derive the equations
\eqref{eqn:V1}-\eqref{eqn:A0}. By taking the $\bz$-average of
\eqref{asym:phi2} and \eqref{asym:A2}, one can get the effective
equations for $U_0(t, \bx)$ and $\bA_0(t, \bx)$. However, in order
to get equations in explicit form, one needs the expressions of
$\rho_1$, $\average{\rho_2}_{\bz}$, $\bJ_0$ and
$\average{\bJ_1}_{\bz}$ in terms of Bloch wave function
$\{\psi_{n,\bk}\}$ or its periodic part $\{u_{n,\bk}\}$. This will
require the following three lemmas. The first lemma states the
property of perturbed density under Hamiltonian perturbation, and
the other two introduce some identities related with Bloch waves.
These identities will be useful in simplifying the expressions of
density and current.

\begin{lemma}\label{lem:denpert}
  We consider the electron dynamics under the perturbed Hamiltonian
  $\tilde{H}=H_0+\sum_{j=1}^J\veps^j
  (V_j+i\bA_{j-1}\cdot\nabla_{\bzeta})$ and denote the density as
\[\tilde{\rho}(t,\bx,\bz)=\rho_0(t,\bx,\bz)+\sum_k\veps^k\tilde{\rho}_k(t,\bx,\bz).\]
Assume the initial condition is
\[\tilde{\rho}(0,\bx,\bz)=\rho_{\gs}(\bz),\] then one
has for any $k$,
\[\langle \tilde{\rho}_k(t,\bx,\bz)\rangle_{\bz}=0.\]
\end{lemma}

\begin{proof}
    Since we choose the Coulomb gauge so that
    $\nabla_{\bzeta}\cdot\bA_{j-1}=0,\;j=1,\cdots,J$, the operator
  $\mathcal{T}\exp\bigl(-i\int_0^t\tilde{H}(\tau)\rd\tau\bigr)$ is unitary, which produces
  \begin{equation*}
    \langle \tilde{\rho}(t,\bx,\bz)
    \rangle_{\bz}=\langle \tilde{\rho}(0,\bx,\bz) \rangle_{\bz}
    =\langle \rho_{\gs}(\bz) \rangle_{\bz}.
  \end{equation*}
  Therefore by \eqref{eqn:rho0_avg} and
  \[\langle\tilde{\rho}(t,\bx,\bz)\rangle_{\bz}=\langle\rho_0(t,\bx,\bz)\rangle_{\bz}
  +\sum_k\veps^k\langle\tilde{\rho}_k(t,\bx,\bz)\rangle_{\bz},\] one
  gets, for any $k$,
  \[\langle \tilde{\rho}_k(t,\bx,\bz)\rangle_{\bz}=0.\]
\end{proof}

\begin{lemma}\label{lem:linear}
  For any $n, m \in \mathbb{Z}_+$ and $\bk, \bp \in \Gamma^{\ast}$, the
following equations hold in the distributional sense.
  \begin{align} & \label{eqn:xi-z} \int_{\RR^3} (\bzeta - \bz)
    \psi_{n,\bk}^{\ast} \psi_{m,\bp} \rd \bzeta =
    \biggl\{\delta_{nm}(-\bz-i\partial_{\bp}) \\
    & \hspace{12em} +\langle u_{n,\bk}|i\nabla_{\bk}|u_{m,\bk}
    \rangle_{L^2(\Gamma)}\biggr\}\delta(\bp-\bk)\abs{\Gamma^*}, \nn \\
    & \int_{\RR^3} v_1(\tau, \bx, \bzeta) \psi_{n,\bk}^{\ast}
    \psi_{m,\bp} \ud\bzeta =\langle
    u_{n,\bk}|v_1(\tau)|u_{m,\bk}\rangle_{L^2(\Gamma)}
    \delta(\bp-\bk)\abs{\Gamma^*},\label{eqn:int_v1}\\
    & \int_{\RR^3} \psi_{n,\bk}^{\ast} (i\nabla_{\bzeta}) \psi_{m,\bp}
    \ud\bzeta =\langle
    u_{n,\bk}|i\nabla_{\bzeta}|u_{m,\bk}\rangle_{L^2(\Gamma)}\delta(\bp-\bk)
    \abs{\Gamma^*}.
    \label{eqn:dzeta}
  \end{align}
\end{lemma}

\begin{proof}
  By the definition of Bloch wave, direct calculation yields
  \begin{equation*}
    \begin{aligned}
      \int_{\RR^3} (\bzeta - \bz) \psi_{n,\bk}^{\ast} \psi_{m,\bp} \rd
      \bzeta & = \int_{\RR^3} (\bzeta - \bz) e^{i\bzeta\cdot(\bp -
        \bk)}
      u_{n,\bk}^{\ast} u_{m,\bp} \ud \bzeta \\
      & = \sum_{\bX_j \in \LL} \int_{\Gamma} (\bzeta + \bX_j - \bz)
      e^{i(\bzeta+\bX_j)\cdot(\bp - \bk)}  u_{n,\bk}^{\ast} u_{m,\bp} \ud \bzeta \\
      & = \sum_{\bX_j \in \LL} \int_{\Gamma} \biggl((-i\partial_{\bp}
      - \bz) e^{i(\bzeta+\bX_j)\cdot(\bp - \bk)}\biggr)
      u_{n,\bk}^{\ast} u_{m,\bp} \ud \bzeta.
    \end{aligned}
  \end{equation*}
  For the right hand side, we have
  \begin{equation}\label{eq:tempeqn1}
    \begin{aligned}
      \sum_{\bX_j \in \LL} \int_{\Gamma} \biggl(-i\partial_{\bp} &
      e^{i(\bzeta+\bX_j)\cdot(\bp - \bk)}\biggr) u_{n,\bk}^{\ast}
      u_{m,\bp} \ud \bzeta  = \\
      & -i \partial_{\bp} \biggl(\sum_{\bX_j \in \LL} \int_{\Gamma}
      e^{i(\bzeta+\bX_j)\cdot(\bp - \bk)} u_{n,\bk}^{\ast}
      u_{m,\bp} \ud \bzeta\biggr) \\
      & \qquad + \sum_{\bX_j \in \LL} \int_{\Gamma}
      e^{i(\bzeta+\bX_j)\cdot(\bp - \bk)}
      u_{n,\bk}^{\ast} (i \partial_{\bp}) u_{m,\bp} \ud \bzeta \\
      = & -i \partial_{\bp} \biggl(\sum_{\bX_j \in \LL} e^{i
        \bX_j\cdot(\bp - \bk)} \int_{\Gamma} e^{i \bzeta\cdot(\bp -
        \bk)}
      u_{n,\bk}^{\ast} u_{m,\bp} \ud \bzeta\biggr) \\
      & \qquad + \sum_{\bX_j \in \LL} e^{i \bX_j\cdot(\bp - \bk)}
      \int_{\Gamma} e^{i \bzeta \cdot(\bp - \bk)}
      u_{n,\bk}^{\ast} (i \partial_{\bp})u_{m,\bp} \ud \bzeta. \\
    \end{aligned}
  \end{equation}

  To further simplify the above expression, we use the Poisson
  summation formula
  \begin{equation}\label{formula:Poisson}
    \sum_{\bX_j \in \LL} e^{i \bX_j\cdot \bk} = \abs{\Gamma^{\ast}}
    \sum_{\bK_j \in \LL^{\ast}} \delta( \bk - \bK_j),
  \end{equation}
  in the distributional sense. Substitute \eqref{formula:Poisson} into
  \eqref{eq:tempeqn1}, we obtain
  \begin{align*}
    \sum_{\bX_j \in \LL} \int_{\Gamma}& \biggl(-i\partial_{\bp}
    e^{i(\bzeta+\bX_j)\cdot(\bp - \bk)}\biggr) u_{n,\bk}^{\ast}
    u_{m,\bp} \ud \bzeta = \\
    & - \abs{\Gamma^{\ast}} i \partial_{\bp} \biggl(\delta(\bp - \bk)
    \int_{\Gamma} e^{i \bzeta\cdot(\bp - \bk)}
    u_{n,\bk}^{\ast} u_{m,\bp} \ud \bzeta\biggr) \\
    & \qquad + \abs{\Gamma^{\ast}} \delta(\bp - \bk) \int_{\Gamma}
    e^{i \bzeta \cdot(\bp - \bk)} u_{n,\bk}^{\ast}
    (i \partial_{\bp}) u_{m,\bp} \ud \bzeta \\
    = & - \abs{\Gamma^{\ast}} i \partial_{\bp} \biggl(\delta(\bp -
    \bk) \int_{\Gamma} e^{i \bzeta\cdot(\bp - \bk)}
    u_{n,\bk}^{\ast} u_{m,\bp} \ud \bzeta\biggr) \\
    & \qquad + \abs{\Gamma^{\ast}} \delta(\bp - \bk) \langle u_{n,\bk}
    \vert i\nabla_{\bk} \vert u_{m,\bk}
    \rangle_{L^2(\Gamma)} \\
    &=- \abs{\Gamma^{\ast}} i \partial_{\bp} \biggl(\delta(\bp - \bk)
    \langle u_{n,\bk}
    \vert u_{m,\bk} \rangle_{L^2(\Gamma)}\biggr)\\
    & \qquad + \abs{\Gamma^{\ast}} \delta(\bp - \bk) \langle u_{n,\bk}
    \vert i\nabla_{\bk} \vert u_{m,\bk}
    \rangle_{L^2(\Gamma)} \\
    &= \abs{\Gamma^{\ast}} (-i \partial_{\bp})\delta(\bp -
    \bk)\delta_{mn} + \abs{\Gamma^{\ast}} \delta(\bp - \bk) \langle
    u_{n,\bk} \vert i\nabla_{\bk} \vert u_{m,\bk}
    \rangle_{L^2(\Gamma)}.
  \end{align*}
  The last equality follows from the orthogonality of $\{u_{n,\bk}\}$
  for each $\bk$.

  Similarly we have
  \begin{equation}\label{eq:term2}
    \begin{aligned}
      \sum_{\bX_j \in \LL} \int_{\Gamma} - \bz
      & e^{i(\bzeta+\bX_j)\cdot(\bp - \bk)} u_{n,\bk}^{\ast} u_{m,\bp} \ud \bzeta \\
      & = - \bz \abs{\Gamma^{\ast}} \delta(\bp - \bk) \int_{\Gamma}
      e^{i
        \bzeta \cdot(\bp - \bk)} u_{n,\bk}^{\ast} u_{m,\bp} \ud \bzeta \\
      & = - \bz \abs{\Gamma^{\ast}} \delta(\bp - \bk) \langle
      u_{n,\bk} \vert u_{m,\bk} \rangle_{L^2(\Gamma)} \\ & = - \bz
      \abs{\Gamma^{\ast}} \delta(\bp - \bk) \delta_{mn},
    \end{aligned}
  \end{equation}
  and
  \begin{equation*}
    \begin{aligned}
      \int_{\RR^3} v_1(\tau,\bx,\bzeta) \psi_{n,\bk}^{\ast}
      \psi_{m,\bp} \rd \bzeta & = \int_{\RR^3} v_1(\tau,\bx,\bzeta)
      e^{i\bzeta\cdot(\bp - \bk)}
      u_{n,\bk}^{\ast} u_{m,\bp} \ud \bzeta \\
      & = \sum_{\bX_j \in \LL} \int_{\Gamma} v_1(\tau,\bx,\bzeta)
      e^{i(\bzeta+\bX_j)\cdot(\bp - \bk)}  u_{n,\bk}^{\ast} u_{m,\bp} \ud \bzeta \\
      & = \langle u_{n,\bk}|v_1(\tau)|u_{m,\bk}\rangle_{L^2(\Gamma)}
      \delta(\bp-\bk)\abs{\Gamma^*},
    \end{aligned}
  \end{equation*}
  where we have used the periodicity of $v_1(\tau,\bx,\bzeta)$ in
  $\bzeta$.

  Hence combining the above equations together yields \eqref{eqn:xi-z},
  and the last equality proves \eqref{eqn:int_v1}. The proof of
  \eqref{eqn:dzeta} is essentially the same as \eqref{eqn:int_v1}
  which we will omit here.
\end{proof}

\begin{lemma}
  For any $n, m \in \mathbb{Z}_+$ and $\bk, \bp \in \Gamma^{\ast}$, we
  have in the distributional sense,
  \begin{equation}\label{eqn:xixi}
    \begin{aligned}
      \int_{\RR^3} (\bzeta-\bz)_{\alpha}(\bzeta-\bz)_{\beta} &
      \psi_{n,\bk}^{\ast} \psi_{m,\bp} \ud\bzeta = \\ &  \delta_{nm}
      (\bz_{\alpha} \bz_{\beta} - \partial_{\bp_{\alpha}}
      \partial_{\bp_{\beta}}) \delta(\bp - \bk) \abs{\Gamma^{\ast}} \\
      & + \langle u_{n,\bk}| i\partial_{\bk_{\beta}} |u_{m,\bk}
      \rangle_{L^2(\Gamma)} (-i \bz_{\alpha}
      -i \partial_{\bp_{\alpha}}) \delta(\bp-\bk) \abs{\Gamma^{\ast}} \\
      & +  \langle u_{n,\bk}| i\partial_{\bk_{\alpha}}| u_{m,\bk}
      \rangle_{L^2(\Gamma)} (-i \bz_{\beta} - i\partial_{\bp_{\beta}})
      \delta(\bp-\bk) \abs{\Gamma^{\ast}} \\
      & - \langle
      u_{n,\bk}| \partial_{\bk_{\alpha}} \partial_{\bk_{\beta}}|
      u_{m,\bk} \rangle_{L^2(\Gamma)} \delta(\bp-\bk)
      \abs{\Gamma^{\ast}}.
    \end{aligned}
  \end{equation}

  \begin{equation} \label{eqn:xiv1}
    \begin{aligned}
      \int_{\RR^3}(\bzeta-\bz)_{\alpha}\partial_{\bx_\alpha} &
      v_1(\tau,\bx,\bzeta)\psi_{n,\bk}^*\psi_{m,\bp}\rd\bzeta = \\
      & \langle
      u_{n,\bk}\rvert \partial_{\bx_\alpha}v_1i\partial_{\bk_{\alpha}} \lvert
      u_{m,\bk}\rangle_{L^2(\Gamma)}\delta(\bp-\bk)\abs{\Gamma^*} \\
      &+\langle u_{n,\bk} \lvert
      \partial_{\bx_\alpha}v_1 \rvert u_{m,\bk}\rangle_{L^2(\Gamma)}
      (-\bz_{\alpha}-i\partial_{\bp_{\alpha}})\delta(\bp-\bk)\abs{\Gamma^*}.
    \end{aligned}
\end{equation}

\begin{equation} \label{eqn:xiA0}
  \begin{aligned}
    \int_{\RR^3}(\bzeta-\bz)_{\alpha}\psi_{n,\bk}^* &
    (i\partial_{\bzeta_{\beta}}) \psi_{m,\bp}\rd\bzeta = \\
    & \langle u_{n,\bk}\rvert
    i\partial_{\bk_{\alpha}}(-\bk_{\beta}+i\partial_{\bzeta_{\beta}})
    \lvert u_{m,\bk}\rangle_{L^2(\Gamma)}\delta(\bp-\bk)\abs{\Gamma^*} \\
    &+\langle u_{n,\bk} \lvert i\partial_{\bzeta_{\beta}} \rvert
    u_{m,\bk}\rangle_{L^2(\Gamma)}
    (-\bz_{\alpha}-i\partial_{\bp_{\alpha}})\delta(\bp-\bk)\abs{\Gamma^*}.
  \end{aligned}
\end{equation}
\end{lemma}

\begin{proof}
  We calculate
  \begin{equation*}
    \begin{aligned}
      \int_{\RR^3} \bzeta_{\alpha}\bzeta_{\beta} \psi_{n,\bk}^{\ast}
      \psi_{m,\bp} \ud \bzeta = & \int_{\RR^3} \bzeta_{\alpha}\bzeta_{\beta}
      u_{n,\bk}^{\ast} u_{m,\bp} e^{i(\bp-\bk)\bzeta} \ud \bzeta \\
      = & - \sum_{\bX_j\in\,\mathbb{L}}\int_{\Gamma} u_{n,\bk}^{\ast}
      u_{m,\bp} \partial_{\bp_{\alpha}}\partial_{\bp_{\beta}}
      e^{i(\bp-\bk)\cdot(\bzeta+\bX_j)} \ud \bzeta.
    \end{aligned}
  \end{equation*}
  The Leibniz rule gives
  \begin{equation*}
    \begin{aligned}
      \int_{\RR^3} \bzeta_{\alpha}\bzeta_{\beta} \psi_{n,\bk}^{\ast}
      & \psi_{m,\bp} \ud \bzeta = \\
      & -\partial_{\bp_{\alpha}}\partial_{\bp_{\beta}} \biggl(
      \int_{\Gamma} u_{n,\bk}^{\ast} u_{m,\bp}
      e^{i(\bp-\bk)\cdot\bzeta}
      \ud\bzeta \sum_{\bX_j\in\mathbb{L}} e^{i(\bp-\bk)\cdot\bX_j}\biggr) \\
      & \qquad +\partial_{\bp_{\alpha}} \biggl( \int_{\Gamma}
      u_{n,\bk}^{\ast} (\partial_{\bp_{\beta}} u_{m,\bp})
      e^{i(\bp-\bk)\cdot\bzeta}
      \ud\bzeta \sum_{\bX_j\in\mathbb{L}} e^{i(\bp-\bk)\cdot\bX_j}\biggr) \\
      & \qquad +\partial_{\bp_{\beta}} \biggl( \int_{\Gamma}
      u_{n,\bk}^{\ast} (\partial_{\bp_{\alpha}} u_{m,\bp})
      e^{i(\bp-\bk)\cdot\bzeta}
      \ud\bzeta \sum_{\bX_j\in\mathbb{L}} e^{i(\bp-\bk)\cdot\bX_j} \biggr)\\
      & \qquad - \int_{\Gamma} u_{n,\bk}^{\ast} (
      \partial_{\bp_{\alpha}} \partial_{\bp_{\beta}} u_{m,\bp})
      e^{i(\bp-\bk)\cdot\bzeta} \ud\bzeta \sum_{\bX_j\in\mathbb{L}}
      e^{i(\bp-\bk)\cdot\bX_j}.
    \end{aligned}
  \end{equation*}
  Therefore applying the Poisson summation formula \eqref{formula:Poisson} yields
  \begin{equation*}
    \begin{aligned}
      \int_{\RR^3} \bzeta_{\alpha}\bzeta_{\beta} \psi_{n,\bk}^{\ast}
      \psi_{m,\bp} \ud \bzeta = & - \delta_{nm} \partial_{\bp_{\alpha}}
      \partial_{\bp_{\beta}} \delta(\bp - \bk) \abs{\Gamma^{\ast}} \\
      & + \langle u_{n,\bk}, \partial_{\bk_{\beta}} u_{m,\bk}
      \rangle_{L^2(\Gamma)} \partial_{\bp_{\alpha}}\delta(\bp-\bk)
      \abs{\Gamma^{\ast}} \\
      & + \langle u_{n,\bk}, \partial_{\bk_{\alpha}} u_{m,\bk}
      \rangle_{L^2(\Gamma)} \partial_{\bp_{\beta}}\delta(\bp-\bk)
      \abs{\Gamma^{\ast}} \\
      & - \langle
      u_{n,\bk}, \partial_{\bk_{\alpha}}\partial_{\bk_{\beta}}
      u_{m,\bk} \rangle_{L^2(\Gamma)}\delta(\bp-\bk)
      \abs{\Gamma^{\ast}},
    \end{aligned}
  \end{equation*}
  and hence using \eqref{eqn:xi-z} and \eqref{eq:term2}, we have
  \eqref{eqn:xixi}.

  The calculations for \eqref{eqn:xiv1} and \eqref{eqn:xiA0} are
  analogous and omitted here.

\end{proof}

\subsubsection{Derivation of the equation \eqref{eqn:rho1_thm}.} By
\eqref{eqn:rho1_old}, \eqref{op:U0} and \eqref{op:U1} the first
order density perturbation reads as
\begin{equation}\label{eq:rho1}
  \begin{aligned}
    \rho_1(t,\bx,\bz) & = \Bigl(\mc{U}^0_{t,0}(\bx) \mathcal{P}
    \mc{U}^1_{0,t}(\bx,\bz)\Bigr)(\bz,\bz) +
    \Bigl(\mc{U}^1_{t,0}(\bx,\bz) \mathcal{P}
    \mc{U}^0_{0,t}(\bx)\Bigr)(\bz,\bz) \\
    & = i e^{-i t H_0} \mathcal{P} \int_0^t e^{i\tau H_0} \delta
    H_1(\tau,\bx,\bz) e^{i(t-\tau)H_0} \ud\tau (\bz,\bz) +
    \text{c.c.},
\end{aligned}
\end{equation}
where we have used the fact that
\begin{equation*}
  \overline{(\mc{U}^1_{t,0}(\bx,\bz) \mathcal{P} \mc{U}^0_{0,t}(\bx))(\bz,\bz)}
  = (\mc{U}^0_{t,0}(\bx) \mathcal{P} \mc{U}^1_{0,t}(\bx,\bz))(\bz,\bz),
\end{equation*}
as a direct consequence of
\begin{equation*}
  (\mc{U}^1_{t,0}(\bx,\bz) \mathcal{P} \mc{U}^0_{0,t}(\bx))^{\ast}
  = \mc{U}^0_{t,0}(\bx) \mathcal{P} \mc{U}^1_{0,t}(\bx,\bz)
\end{equation*}
in the operator sense.

Substitute \eqref{op:dH_1} into \eqref{eq:rho1}, we obtain
\begin{equation}\label{eq:rho1new}
  \begin{aligned}
    \rho_1(t,\bx,\bz) = & -2\Im e^{-itH_0} \mathcal{P} \int_0^t e^{i \tau
      H_0}(\bzeta-\bz)\cdot\nabla_{\bx}U_0(\tau, \bx)e^{i (t-\tau) H_0}
    \rd\tau(\bz, \bz) \\
    & -2\Im e^{-itH_0} \mathcal{P} \int_0^t e^{i \tau H_0} v_1(\tau, \bx, \bzeta)
    e^{i (t-\tau) H_0} \rd\tau(\bz, \bz), \\
    & -2\Im e^{-itH_0} \mathcal{P} \int_0^t e^{i \tau
      H_0}\bA_0(\tau, \bx)\cdot (i\nabla_{\bzeta})e^{i (t-\tau) H_0}
    \rd\tau(\bz, \bz),
  \end{aligned}
\end{equation}
in getting which, we have used
\begin{align*}
  & -2\Im e^{-itH_0} \mathcal{P} \int_0^t e^{i \tau H_0} U_1(\tau, \bx) e^{i
    (t-\tau) H_0} \rd\tau(\bz, \bz) \\
  = & -2\Im e^{-itH_0} \mathcal{P} e^{itH_0}(\bz,\bz) \int_0^t U_1(\tau, \bx)
  \rd\tau \\
  = & - 2\Im \mathcal{P} \int_0^t U_1(\tau, \bx) \ud\tau = 0.
\end{align*}
The first equality above follows from the fact that $U_1(\tau,\bx)$ is
a number as an operator on $L^2_{\bzeta}$.

\begin{prop}The average of $\rho_1$ with respect to
  the microscopic scale vanishes,
  \begin{equation}\label{eqn:rho1_avg}
    \langle \rho_1(t,\bx,\bz)\rangle_{\bz}=0,
  \end{equation}
  which satisfies the second constraint of \eqref{aspt:rho}
  self-consistently.
\end{prop}

\begin{proof}

  In \eqref{eq:rho1new} the first term of the right hand side is an
  odd function in $\bz$, hence when taken the average over $\bz$, it
  gives zero. The second term is the first order density perturbation
  $\tilde{\rho}_1$ if one takes $J=1$ in Lemma~\ref{lem:denpert},
  whose average over $\bz$ is also zero.


\end{proof}

For a more explicit expression for $\rho_1$, we substitute the
spectral representation of operator $H_0$ into \eqref{eq:rho1new},
\begin{equation}\label{spectral:H0}
  H_0 = \sum_n \barint_{\Gamma^{\ast}} E_{n,\bk} \lvert \psi_{n,\bk} \rangle
  \langle  \psi_{n,\bk} \rvert \ud \bk.
\end{equation}
This gives
\begin{equation}\label{expansion:rho1}
  \begin{aligned}
    \rho_1&(t,\bx,\bz) = \\
    & -2 \Im \sum_{n\leq Z} \sum_{m > Z} \int_0^t
    \barint_{(\Gamma^*)^2}\psi_{n,\bk}(\bz)\psi_{m,\bp}^{\ast}(\bz)
    e^{i(E_{m,\bp} - E_{n,\bk})(t-\tau)} \\
    & \hspace{8em} \times \int_{\RR^3} (\bzeta - \bz)
    \psi_{n,\bk}^{\ast}(\bzeta) \psi_{m,\bp}(\bzeta) \rd \bzeta
    \rd\bk\rd\bp  \cdot \nabla_{\bx}U_0(\tau,\bx)\rd\tau\\
    & \quad - 2\Im \sum_{n\leq Z}\sum_{m > Z} \int_0^t
    \barint_{(\Gamma^*)^2}\psi_{n,\bk}(\bz)\psi_{m,\bp}^{\ast}(\bz)
    e^{i(E_{m,\bp} - E_{n,\bk})(t-\tau)} \\
    & \hspace{8em} \times \int_{\RR^3} v_1(\tau, \bx, \bzeta)
    \psi_{n,\bk}^{\ast}(\bzeta) \psi_{m,\bp}(\bzeta)
    \ud\bzeta \rd\bk\rd\bp\rd\tau, \\
    & \quad - 2\Im \sum_{n\leq Z}\sum_{m > Z} \int_0^t
    \barint_{(\Gamma^*)^2}\psi_{n,\bk}(\bz)\psi_{m,\bp}^{\ast}(\bz)
    e^{i(E_{m,\bp} - E_{n,\bk})(t-\tau)} \\
    & \hspace{8em} \times \int_{\RR^3} \psi_{n,\bk}^{\ast}(\bzeta)
    (i\nabla_{\bzeta}) \psi_{m,\bp}(\bzeta)\cdot\bA_0(\tau, \bx)
    \ud\bzeta \rd\bk\rd\bp\rd\tau.
  \end{aligned}
\end{equation}

Substituting \eqref{eqn:xi-z}-\eqref{eqn:dzeta} in
\eqref{expansion:rho1}, we obtain
%
\begin{equation} \label{eqn:rho1}
  \begin{aligned}
    \rho_1(t,\bx,\bz) & =-2\Im\sum_{n\leq Z}\sum_{m>Z} \int_0^t
    \barint_{\Gamma^{\ast}} u_{n,\bk}(\bz) u_{m,\bk}^{\ast}(\bz)
    e^{i\omega_{mn}(\bk)(t-\tau)} \\
    & \hspace{8em} \times \langle u_{n,\bk}|i\nabla_{\bk}|
    u_{m,\bk} \rangle_{L^2(\Gamma)}\rd\bk \cdot\nabla_{\bx}U_0(\tau,\bx)\rd\tau\\
    &\quad -2\Im\sum_{n\leq Z} \sum_{m>Z} \int_0^t
    \barint_{\Gamma^{\ast}} u_{n,\bk}(\bz) u_{m,\bk}^{\ast}(\bz)
    e^{i\omega_{mn}(\bk)(t-\tau)} \\
    & \hspace{8em} \times \langle
    u_{n,\bk}|v_1(\tau,\bx,\bzeta)|u_{m,\bk} \rangle_{L^2(\Gamma)}\rd\bk
    \rd\tau, \\
    &\quad -2\Im\sum_{n\leq Z} \sum_{m>Z} \int_0^t
    \barint_{\Gamma^{\ast}} u_{n,\bk}(\bz) u_{m,\bk}^{\ast}(\bz)
    e^{i\omega_{mn}(\bk)(t-\tau)} \\
    & \hspace{8em} \times \langle
    u_{n,\bk}|i\nabla_{\bzeta}|u_{m,\bk} \rangle_{L^2(\Gamma)}\cdot \bA_0(\tau,\bx)\rd\bk
    \rd\tau.
  \end{aligned}
\end{equation}
This implies \eqref{eqn:rho1_thm}.

\begin{rem}
  From \eqref{eqn:rho1} and using the orthogonality of $\{u_{n,\bk}\}$
for each $\bk$, we once again see that
  \begin{equation*}
    \langle \rho_1(t,\bx,\bz)\rangle_{\bz}=0.
  \end{equation*}
\end{rem}


\subsubsection{Derivation of the equation \eqref{eqn:U0}} By
\eqref{eqn:rho2_old}, \eqref{op:U0} and \eqref{op:U2}, the second
order density perturbation reads as
\begin{align*}
  \rho_2(t,\bx,\bz) =& - 2\Re e^{-itH_0} \mathcal{P} \int_0^t \int_0^{\tau_2}
  e^{i\tau_1 H_0}\delta H_1(\tau_1, \bx, \bz) e^{i(\tau_2-\tau_1)H_0} \\
  & \hspace{8em} \times \delta H_1(\tau_2, \bx, \bz)e^{i(t-\tau_2)H_0}
  \ud\tau_1 \rd\tau_2(\bz,\bz) \\
  & - 2\Im e^{-i t H_0} \mathcal{P} \int_0^t e^{i\tau H_0} \delta
  H_2(\tau,\bx,\bz) e^{i(t-\tau)H_0} \ud\tau (\bz,\bz) \\
  & + \int_0^t e^{-i(t-\tau_1)H_0}\delta H_1(\tau_1,\bx,\bz)
  e^{-i\tau_1 H_0} \rd\tau_1 \mathcal{P} \\
  & \hspace{8em} \times \int_0^t e^{i\tau_2 H_0}\delta
  H_1(\tau_2,\bx,\bz) e^{i(t-\tau_2)H_0}\rd\tau_2 (\bz,\bz).
\end{align*}
Since $\mathcal{P}$ commutates with $\exp(-itH_0)$, we may simplify
the above expression as
\begin{equation}\label{eq:rho2}
\begin{aligned}
  \rho_2&(t,\bx,\bz) = \\
  & - 2\Re \mathcal{P} \int_0^t \int_0^{\tau_2} e^{i
    (\tau_1-t) H_0}\delta H_1(\tau_1, \bx, \bz) e^{i(\tau_2-\tau_1)H_0} \\
  & \hspace{8em} \times \delta H_1(\tau_2, \bx, \bz)e^{i(t-\tau_2)H_0}
  \ud\tau_1 \rd\tau_2(\bz,\bz) \\
  & - 2\Im \mathcal{P} \int_0^t e^{i(\tau-t) H_0} \delta
  H_2(\tau,\bx,\bz) e^{i(t-\tau)H_0} \ud\tau (\bz,\bz) \\
  &+ \int_0^t e^{-i(t-\tau_1)H_0}\delta H_1(\tau_1,\bx,\bz) \mathcal{P} \\
  & \hspace{8em} \times\int_0^t e^{i(\tau_2-\tau_1) H_0}\delta
  H_1(\tau_2,\bx,\bz) e^{i(t-\tau_2)H_0} \rd\tau_2 \rd\tau_1
  (\bz,\bz).
\end{aligned}
\end{equation}

\begin{prop}\label{lem:rho2} The average of $\rho_2(t,\bx,\bz)$ is
given by
  \begin{equation}\label{eqn:rho2}
    \begin{aligned}
      \langle \rho_2&(t,\bx,\bz) \rangle_{\bz} = \\
       & -2\Im\left\langle \mathcal{P} \int_0^t
        e^{i(\tau-t)H_0}\tfrac{1}{2}\bigl((\bzeta-\bz)\cdot\nabla_{\bx}\bigr)^2
        U_0(\tau,\bx) e^{i(t-\tau) H_0}\ud \tau(\bz,\bz)  \right\rangle_{\bz} \\
      & -2\Im \left \langle \mathcal{P} \int_0^t
        e^{i(\tau-t)H_0}(\bzeta-\bz)\cdot\nabla_{\bx}v_1(\tau,\bx,\bzeta)
        e^{i(t-\tau) H_0}\ud \tau(\bz,\bz) \right \rangle_{\bz},\\
        & -2\Im \left \langle \mathcal{P} \int_0^t
        e^{i(\tau-t)H_0}\bigg((\bzeta-\bz)\cdot\nabla_{\bx}\bA_0(\tau,\bx)\bigg)\cdot
        (i\nabla_{\bzeta})
        e^{i(t-\tau) H_0}\ud \tau(\bz,\bz) \right \rangle_{\bz}.
    \end{aligned}
  \end{equation}
\end{prop}

\begin{proof}
  Substituting the expressions of $\delta H_1$ and $\delta H_2$
  \eqref{op:dH_1}-\eqref{op:dH_2} into \eqref{eq:rho2} and taking
  average with respect to $\bz$, one has
  \begin{align*}
    \langle \rho_2&(t,\bx,\bz) \rangle_{\bz} = I_1 + I_2 \\
    & -2\Im\bigg\langle \mathcal{P} \int_0^t
    e^{i(\tau-t)H_0}\tfrac{1}{2}\bigl((\bzeta-\bz)\cdot\nabla_{\bx}\bigr)^2
    U_0(\tau,\bx) e^{i(t-\tau) H_0}\ud \tau(\bz,\bz)
    \bigg\rangle_{\bz}
    \\
    &-2\Im \bigg\langle \mathcal{P} \int_0^t
    e^{i(\tau-t)H_0}(\bzeta-\bz)\cdot\nabla_{\bx}v_1(\tau,\bx,\bzeta)
    e^{i(t-\tau) H_0}\ud \tau(\bz,\bz)
    \bigg\rangle_{\bz} \\
    & -2\Im \left \langle \mathcal{P}\int_0^t
      e^{i(\tau-t)H_0}\bigg((\bzeta-\bz)\cdot\nabla_{\bx}\bA_0(\tau,\bx)\bigg)
      \cdot (i\nabla_{\bzeta}) e^{i(t-\tau) H_0}\ud \tau(\bz,\bz)
    \right \rangle_{\bz},
  \end{align*}
  where $I_1$ and $I_2$ are given by
  \begin{align*}
    I_1=-&2\Re\bigg\langle \mathcal{P}\int_0^t\int_0^{\tau_2}
    e^{iH_0(\tau_1-t)}(\bzeta-\bz)\cdot\nabla_{\bx}U_0(\tau_1)e^{iH_0(\tau_2-\tau_1)}
    \\&\hspace{4em}\times(\bzeta-\bz)\cdot\nabla_{\bx}U_0(\tau_2)e^{iH_0(t-\tau_2)}
    \rd\tau_1\rd\tau_2(\bz,\bz)\bigg\rangle_{\bz} \\
    +& \bigg\langle \int_0^t
    e^{iH_0(\tau_1-t)}(\bzeta-\bz)\cdot\nabla_{\bx}{U_0}(\tau_1)
    \mathcal{P}\\
    & \hspace{4em}\times\int_0^t
    e^{iH_0(\tau_2-\tau_1)}(\bzeta-\bz)\cdot\nabla_{\bx}{U_0}(\tau_2)
    e^{iH_0(t-\tau_2)}
    \rd\tau_2\rd\tau_1(\bz,\bz)\bigg\rangle_{\bz},
  \end{align*}

  \begin{align*}
    I_2=-&2\Re\bigg\langle \mathcal{P}\int_0^t\int_0^{\tau_2}
    e^{iH_0(\tau_1-t)}(\bzeta-\bz)\cdot\nabla_{\bx}U_0(\tau_1)
    e^{iH_0(\tau_2-\tau_1)}\\
    &\qquad\times
    \bigg(v_1(\tau_2)+\bA_0(\tau_2)\cdot(i\nabla_{\bzeta})\bigg)
    e^{iH_0(t-\tau_2)} \rd\tau_1\rd\tau_2(\bz,\bz)\bigg\rangle_{\bz} \\
    -&2\Re\bigg\langle \mathcal{P}\int_0^t\int_0^{\tau_2}
    e^{iH_0(\tau_1-t)}\bigg(v_1(\tau_1)+\bA_0(\tau_1)\cdot(i\nabla_{\bzeta})\bigg)
    e^{iH_0(\tau_2-\tau_1)}\\
    &\qquad\times
    (\bzeta-\bz)\cdot\nabla_{\bx}U_0(\tau_2)e^{iH_0(t-\tau_2)}
    \rd\tau_1\rd\tau_2(\bz,\bz)\bigg\rangle_{\bz}\\
    +& 2\Re\biggl\langle \int_0^t
    e^{iH_0(\tau_1-t)}(\bzeta-\bz)\cdot\nabla_{\bx}{U_0}(\tau_1)
    \mathcal{P} \\
    & \qquad\times\int_0^t
    e^{iH_0(\tau_2-\tau_1)}\biggl(v_1(\tau_2)+\bA_0(\tau_2)\cdot(i\nabla_{\bzeta})
    \biggr) e^{iH_0(t-\tau_2)}
    \rd\tau_2\rd\tau_1(\bz,\bz)\biggr\rangle_{\bz}\;.
\end{align*}
Remark that when calculating
$\langle\rho_2(t,\bx,\bz)\rangle_{\bz}$, we have dropped out the
$\bz$-average of the odd functions in $\bz$ and the second order
density perturbation functions of
$\tilde{H}=H_0+\veps(v_1+U_1+i\bA_0\cdot\nabla_{\bzeta})+\veps^2
(V_2+\tfrac{1}{2}\abs{\bA_0}^2+i\bA_1\cdot\nabla_{\bzeta})$ by Lemma
\ref{lem:denpert}.

We complete the proof by showing that $I_1=0$ and $I_2=0$.

We denote the second term in $I_1$ as
$I_{1,2}=\Re\langle\int_0^t\int_0^t
K(\tau_1,\tau_2)\rd\tau_1\rd\tau_2(\bz,\bz)\rangle_{\bz}$, then
\begin{align*}
I_{1,2} &=\Re\bigg\langle\int_0^t\int_0^{\tau_1}
K(\tau_1,\tau_2)\rd\tau_2\rd\tau_1(\bz,\bz)\bigg\rangle_{\bz} \\
& \qquad\qquad +\Re\bigg\langle\int_0^t\int_0^{\tau_2}
K(\tau_1,\tau_2)\rd\tau_1\rd\tau_2(\bz,\bz)\bigg\rangle_{\bz} \\
& = 2\Re\bigg\langle\int_0^t\int_0^{\tau_2}
K(\tau_1,\tau_2)\rd\tau_1\rd\tau_2(\bz,\bz)\bigg\rangle_{\bz},
\end{align*}
where the last equality is obtained by switching
$\tau_1\leftrightarrow\tau_2$ in the first term of $I_{1,2}$ and
using the fact that $K(\tau_1,\tau_2)=\overline{K(\tau_2,\tau_1)}$.
Therefore $I_1$ could be rewritten as
\begin{equation}\label{eqn:I1}
\begin{aligned} I_1=-&2\Re\bigg\langle
\mathcal{P}\int_0^t\int_0^{\tau_2}
e^{iH_0(\tau_1-t)}(\bzeta-\bz)_{\alpha}e^{iH_0(\tau_2-\tau_1)}
\\&\hspace{8em}\times(\bzeta-\bz)_{\beta}e^{iH_0(t-\tau_2)}
U^{\alpha\beta}_{\tau_1\tau_2}\rd\tau_1\rd\tau_2(\bz,\bz)\bigg\rangle_{\bz} \\
+& 2\Re\bigg\langle \int_0^t\int_0^{\tau_2}
e^{iH_0(\tau_1-t)}(\bzeta-\bz)_{\alpha}\mathcal{P}
e^{iH_0(\tau_2-\tau_1)}\\&\hspace{8em}\times(\bzeta-\bz)_{\beta}e^{iH_0(t-\tau_2)}
U^{\alpha\beta}_{\tau_1\tau_2}\rd\tau_1\rd\tau_2(\bz,\bz)\bigg\rangle_{\bz},
\end{aligned}
\end{equation}
where we have used the short hand notation
$U^{\alpha\beta}_{\tau_1\tau_2}=\partial_{\bx_\alpha}
U_0(\tau_1)\partial_{\bx_\beta}U_0(\tau_2)$.

Substituting spectral representation of $H_0$ \eqref{spectral:H0} into
\eqref{eqn:I1} gives
\begin{align*}
I_1=-&2\Re\sum_{n\leq
Z}\sum_{m\ell}\bigg\langle\int_0^t\int_0^{\tau_2}\barint_{(\Gamma^*)^3}\psi_{n,
\bk}(\bz)\psi_{\ell,\bq}^*(\bz)e^{iE_{n,\bk}(\tau_1-t)}F^{\alpha}_{n,\bk;m,\bp}
\\&\quad\times e^{iE_{m,\bp}(\tau_2-\tau_1)}
F^{\beta}_{m,\bp;\ell,\bq}e^{iE_{\ell,\bq}(t-\tau_2)}U^{\alpha\beta}_{\tau_1\tau_2}
\rd\tau_1\rd\tau_2\rd\bk\rd\bp\rd\bq\bigg\rangle_{\bz} \\
+&2\Re\sum_{n\leq
Z}\sum_{m\ell}\bigg\langle\int_0^t\int_0^{\tau_2}\barint_{(\Gamma^*)^3}\psi_{m,
\bp}(\bz)\psi_{\ell,\bq}^*(\bz)e^{iE_{m,\bp}(\tau_1-t)}F^{\alpha}_{m,\bp;n,\bk}\\&\quad\times
e^{iE_{n,\bk}(\tau_2-\tau_1)} F^{\beta}_{n,\bk;\ell,\bq}
e^{iE_{\ell,\bq}(t-\tau_2)}U^{\alpha\beta}_{\tau_1\tau_2}
\rd\tau_1\rd\tau_2\rd\bk\rd\bp\rd\bq\bigg\rangle_{\bz},
\end{align*}
where
$F^{\alpha}_{n,\bk;m,\bp}=\int_{\RR^3}(\bzeta-\bz)_{\alpha}\psi_{n,\bk}^*(\bzeta)\psi_{m,\bp}(\bzeta)\rd\bzeta$,
and $F^{\alpha}_{m,\bp; n,\bk}$, $F^{\beta}_{m,\bp; \ell,\bq}$,
$F^{\beta}_{n,\bk;\ell,\bq}$ are defined similarly.

We denote $I_1$ as \[I_1=-2\Re\sum_{n\leq
Z}\sum_{m\ell}N_{\bk\bp\bq}^{nml}+2\Re\sum_{n\leq
Z}\sum_{m\ell}N_{\bp\bk\bq}^{mnl},\] then it is easy to see that, by
switching $(m,\bp)$ and $(n,\bk)$,
\[-\Re\sum_{n\leq
Z}\sum_{m\leq Z}\sum_{\ell}N_{\bk\bp\bq}^{nml}+\Re\sum_{n\leq
Z}\sum_{m\leq Z}\sum_{\ell}N_{\bp\bk\bq}^{mnl}=0.\] Therefore
\[I_1=-2\Re\sum_{n\leq
Z}\sum_{m>Z}\sum_{\ell}N_{\bk\bp\bq}^{nml}+2\Re\sum_{n\leq
Z}\sum_{m>Z}\sum_{\ell}N_{\bp\bk\bq}^{mnl}.\] Making use of the
identity \eqref{eqn:xi-z} produces
\begin{align*}
  &F^{\alpha}_{n,\bk;m,\bp}
   = \biggl\{\delta_{nm}(-\bz_{\alpha}-i\partial_{\bp_{\alpha}})+\langle
u_{n,\bk}|i\partial_{\bk_{\alpha}}|u_{m,\bk}
\rangle_{L^2(\Gamma)}\biggr\}\delta(\bp-\bk)\abs{\Gamma^*}, \\
&F^{\beta}_{m,\bp;\ell,\bq}
   = \biggl\{\delta_{m\ell}(-\bz_{\beta}-i\partial_{\bq_{\beta}})+\langle
u_{m,\bp}|i\partial_{\bp_{\beta}}|u_{\ell,\bp}
\rangle_{L^2(\Gamma)}\biggr\}\delta(\bq-\bp)\abs{\Gamma^*}.
\end{align*}

Then by the orthogonality of $\{u_{n,\bk}\}$ for each $\bk$ and
using integration by parts for the variable $\bp$, one could rewrite
$I_1=I_1^{(1)}+I_1^{(2)}$ which are given by
\begin{equation*}
\begin{aligned}
  I_1^{(1)}&= \\
  & -2\Re\sum_{n\leq
    Z}\sum_{m>Z}\sum_{\ell}\bigg\langle\int_0^t\int_0^{\tau_2}
  \barint_{(\Gamma^*)^2}\psi_{n,
    \bk}\psi_{\ell,\bq}^*e^{iE_{n,\bk}(\tau_1-t)}\langle
  u_{n,\bk}|i\partial_{\bk_{\alpha}}|u_{m,\bk} \rangle_{L^2(\Gamma)}\\
  &\qquad \times e^{iE_{m,\bk}(\tau_2-\tau_1)}
  \delta_{m\ell}(-\bz_{\beta}-i\partial_{\bq_{\beta}})\delta(\bq-\bk)
  e^{iE_{\ell,\bq}(t-\tau_2)} U^{\alpha\beta}_{\tau_1\tau_2}\rd\tau_1
  \rd\tau_2\rd\bk\rd\bq\bigg\rangle_{\bz} \\
  & +2\Re\sum_{n\leq
    Z}\sum_{m>Z}\sum_{\ell}\bigg\langle\int_0^t\int_0^{\tau_2}\barint_{(\Gamma^*)^2}
  \psi_{m, \bk}\psi_{\ell,\bq}^*e^{iE_{m,\bk}(\tau_1-t)}\langle
  u_{m,\bk}|i\partial_{\bk_{\alpha}}|u_{n,\bk} \rangle_{L^2(\Gamma)} \\
  &\qquad \times e^{iE_{n,\bk}(\tau_2-\tau_1)}
  \delta_{n\ell}(-\bz_{\beta}-i\partial_{\bq_{\beta}})\delta(\bq-\bk)
  e^{iE_{\ell,\bq}(t-\tau_2)} U^{\alpha\beta}_{\tau_1\tau_2}\rd\tau_1
  \rd\tau_2\rd\bk\rd\bq\bigg\rangle_{\bz}\\
  =& +2\Re\sum_{n\leq
    Z}\sum_{m>Z}\bigg\langle\int_0^t\int_0^{\tau_2}\barint_{\Gamma^*}
  \langle u_{n,\bk}|i\partial_{\bk_{\alpha}}|u_{m,\bk}
  \rangle_{L^2(\Gamma)}\overline{\langle
    u_{n,\bk}|i\partial_{\bk_{\beta}}|u_{m,\bk}
    \rangle}_{L^2(\Gamma)}\\
  &\qquad\times e^{i(E_{m,\bk}-E_{n,\bk})(t-\tau_1)}
  U^{\alpha\beta}_{\tau_1\tau_2}\rd\tau_1\rd\tau_2\rd\bk\bigg\rangle_{\bz}\\
  & -2\Re\sum_{n\leq
    Z}\sum_{m>Z}\bigg\langle\int_0^t\int_0^{\tau_2}\barint_{\Gamma^*}\langle
  u_{m,\bk}|i\partial_{\bk_{\alpha}}|u_{n,\bk}
  \rangle_{L^2(\Gamma)}\overline{\langle
    u_{m,\bk}|i\partial_{\bk_{\beta}}|u_{n,\bk}
    \rangle}_{L^2(\Gamma)}\\
  &\qquad
  e^{i(E_{n,\bk}-E_{m,\bk})(t-\tau_1)}U^{\alpha\beta}_{\tau_1\tau_2}
  \rd\tau_1 \rd\tau_2\rd\bk\bigg\rangle_{\bz},
\end{aligned}
\end{equation*}

\begin{equation*}
\begin{aligned}
I_1^{(2)}=-&2\Re\sum_{n\leq
Z}\sum_{m>Z}\sum_{\ell}\bigg\langle\int_0^t\int_0^{\tau_2}\barint_{\Gamma^*}\psi_{n,
\bk}\psi_{\ell,\bk}^*e^{iE_{n,\bk}(\tau_1-t)}\langle
u_{n,\bk}|i\partial_{\bk_{\alpha}}|u_{m,\bk} \rangle_{L^2(\Gamma)}
\\&\times e^{iE_{m,\bk}(\tau_2-\tau_1)}\langle
u_{m,\bk}|i\partial_{\bk_{\alpha}}|u_{\ell,\bk}\rangle_{L^2(\Gamma)}
e^{iE_{\ell,\bk}(t-\tau_2)} U^{\alpha\beta}_{\tau_1\tau_2}\rd\tau_1
\rd\tau_2\rd\bk\bigg\rangle_{\bz} \\
+&2\Re\sum_{n\leq
Z}\sum_{m>Z}\sum_{\ell}\bigg\langle\int_0^t\int_0^{\tau_2}\barint_{\Gamma^*}\psi_{m,
\bk}\psi_{\ell,\bk}^*e^{iE_{m,\bk}(\tau_1-t)}\langle
u_{m,\bk}|i\partial_{\bk_{\alpha}}|u_{n,\bk} \rangle_{L^2(\Gamma)}
\\&\times e^{iE_{n,\bk}(\tau_2-\tau_1)}
\langle
u_{n,\bk}|i\partial_{\bk_{\alpha}}|u_{\ell,\bk}\rangle_{L^2(\Gamma)}
e^{iE_{\ell,\bk}(t-\tau_2)} U^{\alpha\beta}_{\tau_1\tau_2}\rd\tau_1
\rd\tau_2\rd\bk\bigg\rangle_{\bz}\\
=-&2\Re\sum_{n\leq
Z}\sum_{m>Z}\bigg\langle\int_0^t\int_0^{\tau_2}\barint_{\Gamma^*}
\langle u_{n,\bk}|i\partial_{\bk_{\alpha}}|u_{m,\bk}
\rangle_{L^2(\Gamma)}{\langle
u_{m,\bk}|i\partial_{\bk_{\beta}}|u_{n,\bk}
\rangle}_{L^2(\Gamma)}\\&\qquad\times
e^{i(E_{m,\bk}-E_{n,\bk})(\tau_2-\tau_1)}
 U^{\alpha\beta}_{\tau_1\tau_2}\rd\tau_1\rd\tau_2\rd\bk\bigg\rangle_{\bz}\\
+&2\Re\sum_{n\leq
Z}\sum_{m>Z}\bigg\langle\int_0^t\int_0^{\tau_2}\barint_{\Gamma^*}\langle
u_{m,\bk}|i\partial_{\bk_{\alpha}}|u_{n,\bk}
\rangle_{L^2(\Gamma)}{\langle
u_{n,\bk}|i\partial_{\bk_{\beta}}|u_{m,\bk}
\rangle}_{L^2(\Gamma)}\\&\qquad
e^{i(E_{n,\bk}-E_{m,\bk})(\tau_2-\tau_1)}U^{\alpha\beta}_{\tau_1\tau_2}
\rd\tau_1 \rd\tau_2\rd\bk\bigg\rangle_{\bz}.
\end{aligned}
\end{equation*}

By observing that
\[\overline{\langle
u_{m,\bk}|i\partial_{\bk_{\beta}}|u_{n,\bk}
\rangle}_{L^2(\Gamma)}={\langle
u_{n,\bk}|i\partial_{\bk_{\beta}}|u_{m,\bk} \rangle}_{L^2(\Gamma)},\]
we get $I_1^{(1)}=0$ and $I_1^{(2)}=0$ since one has the same real
part as its complex conjugate.

Therefore $I_1=0$. Similar arguments will show that $I_2=0$ by
making use of the identity \eqref{eqn:int_v1}, and we omit its
details here.
\end{proof}

Substituting the spectral representation of $H_0$
\eqref{spectral:H0} into \eqref{eqn:rho2} gives
\begin{align*}
  \langle \rho_2&(t, \bx,\bz) \rangle_{\bz} =
  \\
  & -2\Im\bigg\langle\sum_{n\leq
    Z}\sum_{m>Z}\int_0^t\barint_{(\Gamma^*)^2}\psi_{n,\bk}(\bz)\psi_{m,\bp}^*(\bz)
  e^{i(E_{m,\bp}-E_{n,\bk})(t-\tau)}
  \\&\qquad\times\frac{1}{2}\int_{\RR^3}(\bzeta-\bz)_{\alpha}(\bzeta-\bz)_{\beta}
  \psi_{n,\bk}^*(\bzeta)\psi_{m,\bp} \rd\bzeta\rd\bk\rd\bp
  \partial_{\bx_\alpha}\partial_{\bx_\beta}U_0(\tau,\bx)\rd\tau\bigg\rangle_{\bz}
  \\ & -2\Im\bigg\langle\sum_{n\leq
    Z}\sum_{m>Z}\int_0^t\barint_{(\Gamma^*)^2}\psi_{n,\bk}(\bz)
  \psi_{m,\bp}^*(\bz)e^{i(E_{m,\bp}-E_{n,\bk})(t-\tau)}
  \\&\qquad\times\int_{\RR^3}(\bzeta-\bz)_{\alpha}\partial_{\bx_\alpha}
  v_1(\tau,\bx,\bzeta)
  \psi_{n,\bk}^*(\bzeta)\psi_{m,\bp}\rd\bzeta\rd\bk\rd\bp\rd\tau\bigg\rangle_{\bz}.
\end{align*}

By \eqref{eqn:xixi} and \eqref{eqn:xiv1} and using the integration
by parts with respect to $\bp$, we could simplify the above equality
to be
\begin{align*}
  \langle \rho_2& (t,\bx,\bz) \rangle_{\bz}= \\
  & 2\Im \sum_{n\leq Z}\sum_{m>Z}\int_0^t\barint_{\Gamma^*}
  e^{i\omega_{mn}(\bk)(t-\tau)} \\& \qquad\times\overline{\langle
    u_{n,\bk}|i\partial_{\bk_\alpha}|u_{m,\bk}
    \rangle}_{L^2(\Gamma)}\langle
  u_{n,\bk}|i\partial_{\bk_\beta}|u_{m,\bk}
  \rangle_{L^2(\Gamma)}\partial_{\bx_\alpha}\partial_{\bx_\beta}U_0(\tau)\rd\bk\rd\tau
  \\&+2\Im \sum_{n\leq Z} \sum_{m>Z} \int_0^t\barint_{\Gamma^*}
  e^{i\omega_{mn}(\bk)(t-\tau)}\\& \qquad\times\overline{\langle
    u_{n,\bk}|i\partial_{\bk_\alpha}|u_{m,\bk} \rangle}_{L^2(\Gamma)}
  {\langle u_{n,\bk}|\partial_{\bx_\alpha} v_1(\tau)|u_{m,\bk}}
  \rangle_{L^2(\Gamma)}\rd\bk\rd\tau, \\&+2\Im \sum_{n\leq Z}
  \sum_{m>Z} \int_0^t\barint_{\Gamma^*}
  e^{i\omega_{mn}(\bk)(t-\tau)}\\& \qquad\times\overline{\langle
    u_{n,\bk}|i\partial_{\bk_\alpha}|u_{m,\bk} \rangle}_{L^2(\Gamma)}
  {\langle u_{n,\bk}|i\partial_{\bzeta_{\beta}}|u_{m,\bk}}
  \rangle_{L^2(\Gamma)}
    \partial_{\bx_{\alpha}}(\bA_0)_{\beta}\rd\bk\rd\tau.
\end{align*}
Therefore taking the $\bz$-average of \eqref{asym:phi2} produces
\begin{multline*}
  -{\delta_{\alpha\beta}}\partial_{\bx_\alpha}\partial_{\bx_\beta}U_0
  =\langle\rho_2(t,\bx,\bz) \rangle_{\bz}+\rho_{\ext}(t,\bx)\\
  = P_{\alpha\beta}(\partial_{\bx_\alpha}\partial_{\bx_\beta}
  U_0)+Q_{\alpha}\left(\partial_{\bx_\alpha}
    v_1\right)+R_{\alpha\beta}(\partial_{\bx_{\alpha}(\bA_0)_{\beta}}
  +\rho_{\ext}(t,\bx).
\end{multline*}
This proves \eqref{eqn:U0}.

\subsubsection{Derivation of the equation \eqref{eqn:A0}.}
Substituting the spectral representation of $H_0$
\eqref{spectral:H0} into \eqref{eqn:J0_old} yields
\begin{equation}\label{eqn:J0_new}
  \bJ_0=\sum_{n\leq Z}\barint_{\Gamma^*}\Im
  \psi_{n,\bk}\nabla_{\bzeta}\psi_{n,\bk}\rd\bk.
\end{equation}

\begin{prop} We have \[\average{\bJ_0}_{\bz}=0,\] which satisfies the
  third constraint in \eqref{aspt:rho} self-consistently.
\end{prop}
\begin{proof}
  By definition of the Bloch decomposition, we have
  \begin{equation}\label{eqn:u_nk}
    \tilde{H_0}u_{n,\bk}=\biggl(\frac{1}{2}(-i\nabla_{\bz}
    +\bk)^2+v_0(\bz)\biggr)
    u_{n,\bk}(\bz)=E_{n,\bk}u_{n,\bk}(\bz).
  \end{equation}

  Differentiating \eqref{eqn:u_nk} with respect to $\bk$ gives
  \[(-i\nabla_{\bzeta}+\bk)u_{n,\bk}+\tilde{H_0}\nabla_{\bk}u_{n,\bk}=
  \nabla_{\bk}E_{n,\bk}u_{n,\bk}+E_{n,\bk}\nabla_{\bk}u_{n,\bk}.\]

  Since $\tilde{H_0}$ is a self-adjoint operator, the above equation
  taken the inner product with $u_{n,\bk}$ yields
  \[\average{u_{n,\bk}\lvert -i\nabla_{\bzeta}+\bk \rvert
    u_{n,\bk}}_{L^2(\Gamma)}=\nabla_{\bk}E_{n,\bk}.\]

  Therefore by \eqref{eqn:J0_new},
  \[\average{\bJ_0}_{\bz}=\sum_{n\leq Z}\barint_{\Gamma^*}
  \average{u_{n,\bk}\lvert -i\nabla_{\bzeta}+\bk \rvert
    u_{n,\bk}}_{L^2(\Gamma)} \rd\bk=\sum_{n\leq
    Z}\barint_{\Gamma^*}\nabla_{\bk}E_{n,\bk}\rd\bk=0,\] where the
  last equality is due to the periodicity.
\end{proof}

Similar to the derivation of \eqref{eqn:rho1}, by making use of
Lemma~\ref{lem:linear} and \eqref{aspt:rho}, direct calculations
from \eqref{eqn:J1_old} give the following proposition.

\begin{prop} The average of $\bJ_1(t,\bx,\bz)$ is given by
  \begin{align}\label{eqn:J1}\average{\bJ_1}_{\bz}=
    &2\Im\sum_{n\leq Z}\sum_{m>Z}\int_0^t\barint_{\Gamma^*}
    \overline{\bra{u_{n,\bk}}i\nabla_{\bzeta}\ket{u_{m,\bk}}}_{L^2(\Gamma)}
    e^{i\omega_{mn}(\bk)(t-\tau)}
    \\&\times\biggl(\bra{u_{n,\bk}}i\nabla_{\bk}\ket{u_{m,\bk}}_{L^2(\Gamma)}
    \cdot\nabla_{\bx}U_0(\tau)
    +\bra{u_{n,\bk}}v_1(\tau)\ket{u_{m,\bk}}_{L^2(\Gamma)}\nn\\&\qquad\qquad
    +\bra{u_{m,\bk}}i\nabla_{\bzeta}\ket{u_{n,\bk}}_{L^2(\Gamma)}
    \cdot\bA_0(\tau)
    \biggr)\rd\bk\rd\tau-\bA_0\average{\rho_0}_{\bz}.\nn
  \end{align}
\end{prop}

Then \eqref{eqn:J1} implies \eqref{eqn:A0} by taking the $\bz$-average
of \eqref{asym:A2}.

\section{Effective equations in the frequency domain}\label{sec:freq}

We now derive the effective equations in frequency domain.  We start
with the following proposition of the Fourier transform.

\begin{prop}\label{prop:fourier}
  Define the function $h(t)=\int_0^t h_1(t-\tau)\cdot h_2(\tau)\rd
  \tau$, then
  \[\hat{h}=\hat{h_1H_v}\cdot\hat{h_2H_v},\]
  where $H_v(t)$ is the Heaviside function of $t$,
  \begin{equation*}
  H_v(t)=\begin{cases} 1\qquad t\geq 0, \\ 0 \qquad \hbox{otherwise}.
  \end{cases}
\end{equation*}
\end{prop}

Without loss of generality, we assume that $U_0(t)$ and $\bA_0(t)$
vanish for $t<0$. By taking the Fourier transform of
\eqref{eqn:V1}-\eqref{eqn:A0} and using
Proposition~\ref{prop:fourier}, we have
\begin{align}\label{eqn:U0hat}
  & -\bigl(\delta_{\alpha\beta}+A_{\alpha\beta}\bigr)\partial_{\bx_{\alpha}}
  \partial_{\bx_{\beta}}\hat{U_0} =
  B_{\alpha\beta}\bigl(\partial_{\bx_{\alpha}}(\hat{\bA_0})_{\beta}\bigr)
  +\hat{\rho_{\ext}}, \\ \label{eqn:A0hat} &
  -\omega^2(\hat{\bA_0})_{\alpha}-\Delta_{\bx}(\hat{\bA_0})_{\alpha}
  +(-i\omega)\partial_{\bx_{\alpha}}\hat{U_0}
  =C_{\alpha\beta}\partial_{\bx_{\beta}}\hat{U_0}+D_{\alpha\beta}(\hat{\bA_0})_{\beta}
  +\hat{\bJ_{\ext}}.
\end{align}

The coefficients are given by
\begin{align*}
  & A_{\alpha\beta}=
  \hat{P}_{\alpha\beta}-\average{\hat{\bbf}^*_{\alpha}\mathcal{V}
    (I-\hat{\chi}_{\omega}\mathcal{V})^{-1}
    \hat{\bbf}_{\beta}}_{\bz}, \\
  &
  B_{\alpha\beta}=\hat{R}_{\alpha\beta}-\average{\hat{\bbf}^*_{\alpha}
    \mathcal{V}(I-\hat{\chi}_{\omega}\mathcal{V})^{-1}
    \hat{\bbg}_{\beta}}_{\bz}, \\
  &
  C_{\alpha\beta}=\hat{M}_{\alpha\beta}-\average{\hat{\bbg}^*_{\alpha}
    \mathcal{V}(I-\hat{\chi}_{\omega}\mathcal{V})^{-1}
    \hat{\bbf}_{\beta}}_{\bz}, \\
  &
  D_{\alpha\beta}=\hat{N}_{\alpha\beta}-\average{\hat{\bbg}^*_{\alpha}
    \mathcal{V}(I-\hat{\chi}_{\omega}\mathcal{V})^{-1}
    \hat{\bbg}_{\beta}}_{\bz}-\delta_{\alpha\beta}\average{\rho_0}_{\bz},
\end{align*}
where
\begin{align*}
  \hat{\chi}_{\omega}\hat{v_1}(\omega)=&-\sum_{n\leq
    Z}\sum_{m>Z}\barint_{\Gamma^*}\frac{1}{\omega + \omega_{mn}(\bk)}
  u_{n,\bk} u_{m,\bk}^*\langle
  u_{n,\bk}|\hat{v_1}(\omega)|u_{m,\bk} \rangle_{L^2(\Gamma)}\rd\bk \\
  &+\sum_{n\leq Z}\sum_{m>Z}\barint_{\Gamma^*}\frac{1}{\omega -
    \omega_{mn}(\bk)} u_{n,\bk}^* u_{m,\bk}\overline{\langle
    u_{n,\bk}|\hat{v_1}(\omega)|u_{m,\bk} \rangle}_{L^2(\Gamma)}\rd\bk, \\
  \hat{\bbf}(\omega)=&-\sum_{n\leq
    Z}\sum_{m>Z}\barint_{\Gamma^*}\frac{1}{\omega + \omega_{mn}(\bk)}
  u_{n,\bk} u_{m,\bk}^*\langle
  u_{n,\bk}|i\nabla_{\bk}|u_{m,\bk} \rangle_{L^2(\Gamma)}\rd\bk \\
  &+\sum_{n\leq Z}\sum_{m>Z}\barint_{\Gamma^*}\frac{1}{\omega -
    \omega_{mn}(\bk)} u_{n,\bk}^* u_{m,\bk}\overline{\langle
    u_{n,\bk}|i\nabla_{\bk}|u_{m,\bk} \rangle}_{L^2(\Gamma)}\rd\bk, \\
  \hat{\bbg}(\omega)=&-\sum_{n\leq
    Z}\sum_{m>Z}\barint_{\Gamma^*}\frac{1}{\omega + \omega_{mn}(\bk)}
  u_{n,\bk} u_{m,\bk}^*\langle
  u_{n,\bk}|i\nabla_{\bzeta}|u_{m,\bk} \rangle_{L^2(\Gamma)}\rd\bk \\
  &+\sum_{n\leq Z}\sum_{m>Z}\barint_{\Gamma^*}\frac{1}{\omega -
    \omega_{mn}(\bk)} u_{n,\bk}^* u_{m,\bk}\overline{\langle
    u_{n,\bk}|i\nabla_{\bzeta}|u_{m,\bk} \rangle}_{L^2(\Gamma)}\rd\bk,
\end{align*}
and
\begin{align*}
  \hat{P}_{\alpha\beta}(\omega)=&\sum_{n\leq
    Z}\sum_{m>Z}\barint_{\Gamma^*}\frac{1}{\omega+\omega_{mn}(\bk)}
  \overline{\bra{u_{n,\bk}}i\partial_{\bk_{\alpha}}\ket{u_{m,\bk}}}
  \bra{u_{n,\bk}}i\partial_{\bk_{\beta}}\ket{u_{m,\bk}}\rd\bk \\
  & -\sum_{n\leq
    Z}\sum_{m>Z}\barint_{\Gamma^*}\frac{1}{\omega-\omega_{mn}(\bk)}
  \bra{u_{n,\bk}}i\partial_{\bk_{\alpha}}\ket{u_{m,\bk}}
  \overline{\bra{u_{n,\bk}}i\partial_{\bk_{\beta}}\ket{u_{m,\bk}}}\rd\bk,
\end{align*}
\begin{align*}
  \hat{R}_{\alpha\beta}(\omega)=&\sum_{n\leq
    Z}\sum_{m>Z}\barint_{\Gamma^*}\frac{1}{\omega+\omega_{mn}(\bk)}
  \overline{\bra{u_{n,\bk}}i\partial_{\bk_{\alpha}}\ket{u_{m,\bk}}}
  \bra{u_{n,\bk}}i\partial_{\bzeta_{\beta}}\ket{u_{m,\bk}}\rd\bk \\
  & -\sum_{n\leq
    Z}\sum_{m>Z}\barint_{\Gamma^*}\frac{1}{\omega-\omega_{mn}(\bk)}
  \bra{u_{n,\bk}}i\partial_{\bk_{\alpha}}\ket{u_{m,\bk}}
  \overline{\bra{u_{n,\bk}}i\partial_{\bzeta_{\beta}}\ket{u_{m,\bk}}}\rd\bk,
\end{align*}
\begin{align*}
  \hat{M}_{\alpha\beta}(\omega)=&\sum_{n\leq
    Z}\sum_{m>Z}\barint_{\Gamma^*}\frac{1}{\omega+\omega_{mn}(\bk)}
  \overline{\bra{u_{n,\bk}}i\partial_{\bzeta_{\alpha}}\ket{u_{m,\bk}}}
  \bra{u_{n,\bk}}i\partial_{\bk_{\beta}}\ket{u_{m,\bk}}\rd\bk \\
  & -\sum_{n\leq
    Z}\sum_{m>Z}\barint_{\Gamma^*}\frac{1}{\omega-\omega_{mn}(\bk)}
  \bra{u_{n,\bk}}i\partial_{\bzeta_{\alpha}}\ket{u_{m,\bk}}
  \overline{\bra{u_{n,\bk}}i\partial_{\bk_{\beta}}\ket{u_{m,\bk}}}\rd\bk,
\end{align*}
\begin{align*}
  \hat{N}_{\alpha\beta}(\omega)=&\sum_{n\leq
    Z}\sum_{m>Z}\barint_{\Gamma^*}\frac{1}{\omega+\omega_{mn}(\bk)}
  \overline{\bra{u_{n,\bk}}i\partial_{\bzeta_{\alpha}}\ket{u_{m,\bk}}}
  \bra{u_{n,\bk}}i\partial_{\bzeta_{\beta}}\ket{u_{m,\bk}}\rd\bk \\
  & -\sum_{n\leq
    Z}\sum_{m>Z}\barint_{\Gamma^*}\frac{1}{\omega-\omega_{mn}(\bk)}
  \bra{u_{n,\bk}}i\partial_{\bzeta_{\alpha}}\ket{u_{m,\bk}}
  \overline{\bra{u_{n,\bk}}i\partial_{\bzeta_{\beta}}\ket{u_{m,\bk}}}\rd\bk.
\end{align*}

We need the following proposition to further simplify the equations.
\begin{prop}\label{prop:der}
\[\average{u_{n,\bk}\lvert i\nabla_{\bzeta}\rvert u_{m,\bk}}_{L^2(\Gamma)}=
i\omega_{mn}(\bk)\average{u_{n,\bk}\lvert i\nabla_{\bk}\rvert
u_{m,\bk}}_{L^2(\Gamma)}.\]
\end{prop}
\begin{proof}
Similar to \eqref{eqn:u_nk} one has
\begin{equation*}\tilde{H_0}u_{m,\bk}=\biggl(\frac{1}{2}(-i\nabla_{\bz} +\bk)^2+v_0(\bz)\biggr)
u_{m,\bk}(\bz)=E_{m,\bk}u_{m,\bk}(\bz).\end{equation*}

Differentiating it with respect to $\bk$ gives
\[(-i\nabla_{\bzeta}+\bk)u_{m,\bk}+\tilde{H_0}\nabla_{\bk}u_{m,\bk}=
\nabla_{\bk}E_{m,\bk}u_{m,\bk}+E_{m,\bk}\nabla_{\bk}u_{m,\bk}.\]

Since $\tilde{H_0}$ is a self-adjoint operator, the above equation
taken the inner product with $u_{n,\bk}$ produces
\[\average{u_{n,\bk}\lvert -i\nabla_{\bzeta}\rvert u_{m,\bk}}_{L^2(\Gamma)}=(E_{m,\bk}-E_{n,\bk})
\average{u_{n,\bk}\lvert \nabla_{\bk}\rvert
u_{m,\bk}}_{L^2(\Gamma)},\] which implies the conclusion.
\end{proof}

\begin{lemma}
\begin{align}\label{constraint:1}
& 2\Im \sum_{n\leq
Z}\sum_{m>Z}\barint_{\Gamma^*}u_{n,\bk}(\bz)u_{m,\bk}^*(\bz)\average{u_{n,\bk}\lvert
i\nabla_{\bk}\rvert u_{m,\bk}}_{L^2(\Gamma)}\rd\bk=0, \\
& \label{constraint:2} -2\Im\sum_{n\leq
Z}\sum_{m>Z}\barint_{\Gamma^*}\overline{\average{u_{n,\bk}\lvert
i\partial_{\bzeta_{\alpha}}\rvert
u_{m,\bk}}}_{L^2(\Gamma)} \\
& \hspace{10em} \times \average{u_{n,\bk}\lvert
i\partial_{\bk_{\beta}}\rvert
u_{m,\bk}}_{L^2(\Gamma)}\rd\bk=\average{\rho_0}_{\bz}\delta_{\alpha\beta}.\nn
\end{align}
\end{lemma}
\begin{proof}
Since adding any constant vector to $\bA_0$ will not change the
system \eqref{eqn:Schr}-\eqref{eqn:VecPot}, the values of $\rho_1$
and $\average{\bJ_1}_{\bz}$ remain the same under the transform
$\bA_0\rightarrow \bA_0 + \boldsymbol{C_{v}}$ where
$\boldsymbol{C_{v}}$ is an arbitrary constant vector.

Note that we have assumed $\bA_0(t)=0$ for $t<0$, then
\eqref{eqn:rho1} implies that
\begin{multline*}
  2\Im \sum_{n\leq
    Z}\sum_{m>Z}\int_{-\infty}^t\barint_{\Gamma^*}u_{n,\bk}(\bz)u_{m,\bk}^*(\bz)
  e^{i\omega_{mn}(\bk)(t-\tau)} \\\times \average{u_{n,\bk} \lvert
    i\nabla_{\bk}\rvert u_{m,\bk}}_{L^2(\Gamma)}\rd\bk\rd\tau=0,
\end{multline*}
which gives \eqref{constraint:1} by making use of
Proposition~\ref{prop:der}.

Similarly \eqref{eqn:J1} implies
\begin{multline*}-2\Im\sum_{n\leq
Z}\sum_{m>Z}\int_{-\infty}^t\barint_{\Gamma^*}e^{i\omega_{mn}(\bk)(t-\tau)}\\
\times\overline{\average{u_{n,\bk}\lvert
i\partial_{\bzeta_{\alpha}}\rvert
u_{m,\bk}}}_{L^2(\Gamma)}\average{u_{n,\bk}\lvert
i\partial_{\bk_{\beta}}\rvert
u_{m,\bk}}_{L^2(\Gamma)}\rd\bk\rd\tau=\average{\rho_0}_{\bz}\delta_{\alpha\beta},\end{multline*}
which produces \eqref{constraint:2} by making use of
Proposition~\ref{prop:der}.
\end{proof}

\begin{lemma}\label{lem:coeff}
\begin{align*} & \hat{\bbg}(\omega)=(-i\omega)\hat{\bbf}(\omega), \quad
\hat{N}_{\alpha\beta}(\omega)-\average{\rho_0}_{\bz}\delta_{\alpha\beta}=(-i\omega)\hat{M}_{\alpha\beta}(\omega),
\\ & \hat{M}_{\alpha\beta}(\omega)=-\hat{R}_{\alpha\beta}(\omega),\quad \hat{R}_{\alpha\beta}(\omega)
=(-i\omega)\bigl(\hat{P}_{\alpha\beta}(\omega)-P^r_{\alpha\beta}\bigr),\end{align*}
where
\[P^r_{\alpha\beta}=\frac{2i}{\omega}\Im\sum_{n\leq Z}\sum_{m>Z}\barint_{\Gamma^*}
\overline{\bra{u_{n,\bk}}i\partial_{\bk_{\alpha}}\ket{u_{m,\bk}}}
\bra{u_{n,\bk}}i\partial_{\bk_{\beta}}\ket{u_{m,\bk}}\rd\bk,\] which
satisfies $P^{r}_{\alpha\beta}=-P^r_{\beta\alpha}$.
\end{lemma}
\begin{proof}
Observe that
\[\frac{i\omega_{mn}(\bk)}{\omega+\omega_{mn}(\bk)}=i+\frac{-i\omega}{\omega+\omega_{mn}(\bk)},
\quad
\frac{-i\omega_{mn}(\bk)}{\omega-\omega_{mn}(\bk)}=i+\frac{-i\omega}{\omega-\omega_{mn}(\bk)}.
\]

Then it is easy to see that Proposition~\ref{prop:der} along with
\eqref{constraint:1} implies
$\hat{\bbg}(\omega)=(-i\omega)\hat{\bbf}(\omega)$, and
Proposition~\ref{prop:der} along with \eqref{constraint:2} implies
$\hat{N}_{\alpha\beta}(\omega)-\average{\rho_0}_{\bz}\delta_{\alpha\beta}
=(-i\omega)\hat{M}_{\alpha\beta}(\omega)$.

Moreover, Proposition~\ref{prop:der} also implies
\begin{align*}
\hat{M}_{\alpha\beta}(\omega)=&-\hat{R}_{\alpha\beta}(\omega),\\
\hat{R}_{\alpha\beta}(\omega)=&(-i\omega)\hat{P}_{\alpha\beta}(\omega)\\
& +i\sum_{n\leq Z}\sum_{m>Z}\barint_{\Gamma^*}
\overline{\bra{u_{n,\bk}}i\partial_{\bk_{\alpha}}\ket{u_{m,\bk}}}
\bra{u_{n,\bk}}i\partial_{\bk_{\beta}}\ket{u_{m,\bk}}\rd\bk \\
&-i\sum_{n\leq Z}\sum_{m>Z}\barint_{\Gamma^*}
\bra{u_{n,\bk}}i\partial_{\bk_{\alpha}}\ket{u_{m,\bk}}
\overline{\bra{u_{n,\bk}}i\partial_{\bk_{\beta}}\ket{u_{m,\bk}}}\rd\bk
\\ =&(-i\omega)\bigl(\hat{P}_{\alpha\beta}(\omega)-P^r_{\alpha\beta}\bigr).
\end{align*}
\end{proof}

Note that
\[
\partial_{\bx_{\alpha}}\partial_{\bx_{\beta}}\hat{U_0}
=\partial_{\bx_{\beta}}\partial_{\bx_{\alpha}}\hat{U_0}, \qquad
P^{r}_{\alpha\beta}=-P^r_{\beta\alpha},
\]
one knows that the equation
\eqref{eqn:U0hat} will remain the same if we redefine
\[
A_{\alpha\beta}=\hat{P}_{\alpha\beta}-P^r_{\alpha\beta}(\omega)
-\average{\hat{\bbf}^*_{\alpha}\mathcal{V}(I-\hat{\chi}_{\omega}\mathcal{V})^{-1}
  \hat{\bbf}_{\beta}}_{\bz}.
\]
Then Lemma~\ref{lem:coeff} implies
\[
B_{\alpha\beta}=(-i\omega)A_{\alpha\beta},\quad
D_{\alpha\beta}=(-i\omega)C_{\alpha\beta}, \quad
C_{\alpha\beta}=-B_{\alpha\beta}.
\]

By defining $\hat{\bE}=-\nabla_{\bx}\hat{U_0}+i\omega \hat{\bA_0},\;
\hat{\bB}=\nabla_{\bx}\times\hat{\bA_0}$, the equations
\eqref{eqn:U0hat}-\eqref{eqn:A0hat} along with
$\nabla_{\bx}\cdot\hat{\bA_0}=0$ produce \eqref{eqn:E}-\eqref{eqn:B}.
This completes the derivation of the main result in
Section~\ref{sec:mainresult}.

\section{Conclusion} \label{sec:conclusion}

One unsatisfactory aspect of this work is that it is limited to
short time scales.  In fact, the behavior at longer time scales
is still very much of a mystery, even from the viewpoint of formal
asymptotics.  This is very unsettling.  The main technical difficulty
is the lack of local charge neutrality and the huge potential
generated as a result.

There are other important issues that remain.
These include the inclusion of spin, the interaction with lattice
dynamics, defects, instabilities, etc.

\bibliographystyle{amsplain}
\bibliography{tddft}

\end{document}